\documentstyle[12pt]{article}

\newcounter{eq}
\newcounter{sc}

\def\overleftrightarrow#1{\vbox{\ialign{##\crcr
 $\leftrightarrow$\crcr\noalign{\kern-1pt\nointerlineskip}
 $\hfil\displaystyle{#1}\hfil$\crcr}}}

\newlength{\minitwocolumn}
\begin{document}

\begin{flushright}
DPUR/TH/73\\
January, 2022\\
\end{flushright}
\vspace{20pt}

\pagestyle{empty}
\baselineskip15pt

\begin{center}
{\large\bf  Quantum Scale Invariant Gravity with de Donder Gauge
\vskip 1mm }

\vspace{20mm}

Ichiro Oda\footnote{
           E-mail address:\ ioda@sci.u-ryukyu.ac.jp
                  }

\vspace{10mm}
           Department of Physics, Faculty of Science, University of the 
           Ryukyus,\\
           Nishihara, Okinawa 903-0213, Japan\\

\end{center}


\vspace{10mm}
\begin{abstract}

We perform the manifestly covariant quantization of a scale invariant gravity with a scalar field, which is 
equivalent to the well-known Brans-Dicke gravity via a field redefinition of the scalar field, in the de Donder 
gauge condition (or harmonic gauge condition) for general coordinate invariance. First, without specifying 
the expression of a gravitational theory, we write down various equal-time (anti-)commutation relations (ETCRs), 
in particular, those involving the Nakanishi-Lautrup field, the FP ghost, and the FP antighost only on the basis 
of the de Donder gauge condition. It is shown that choral symmetry, which is a Poincar${\rm{\acute{e}}}$-like 
$IOSp(8|8)$ supersymmetry, can be derived from such a general action with the de Donder gauge.
Next, taking the scale invariant gravity with a scalar field as a classical theory, we derive the ETCRs for the
gravitational sector involving the metric tensor and scalar fields. Moreover, we account for how scale symmetry is 
spontaneously broken in quantum gravity, thereby showing that the dilaton is a massless Nambu-Goldstone
particle.

\end{abstract}

\newpage
\pagestyle{plain}
\pagenumbering{arabic}


\section{Introduction}

A residual symmetry which is left behind after taking a certain gauge-fixing condition for the gauge invariance, 
has thus far played an important role in quantum field theory.  For instance, in string theory, conformal symmetry 
on the world sheet can be thought as the fundamental symmetry in perturbative regime where strings are weakly 
interacting \cite{Polchinski}. The conformal symmetry is a typical residual symmetry, which is left in a theory 
after taking the conformal gauge for the world-sheet diffeomorphism (or general coordinate transformation (GCT)) 
and the Weyl symmetry (or a local scale transformation) \cite{Zumino}.

Recently, we have elucidated various aspects of such residual symmetries existing in some gravitational theories. 
In particular, in the most recent study, we have shown that using the simplest scalar-tensor gravity \cite{Fujii}
the restricted Weyl symmetry (RWS) and general coordinate invariance generate conformal symmetry in four dimensions in a flat 
Minkowski background \cite{Oda-R}-\cite{Oda-RWS}.  
   
Also about Einstein gravity, in a pioneering work by Nakanishi \cite{Nakanishi, N-O-text}, on the basis of the Einstein-Hilbert 
action in the de Donder gauge (harmonic gauge) for general coordinate invariance, it has been shown that there remains 
a huge residual symmetry, which is a Poincar$\rm{\acute{e}}$-like $ISOp(8|8)$ supersymmetry, called ``choral symmetry'',  
including the BRST symmetry and $GL(4)$ symmetry etc. It is of interest that in this formulation the graviton can be 
identified with a Nambu-Goldstone particle associated with spontaneous symmetry breakdown of $GL(4)$ symmetry 
to the Lorentz symmetry $SO(1, 3)$, thereby proving the exact masslessness of the graviton in a nonperturbative 
manner \cite{NO}. 

However, in this formulation, the Einstein equation was critically used in obtaining the equal-time commutation relation
$[ b_\mu, \dot b_\nu^\prime ] = i \tilde f ( \partial_\mu b_\nu + \partial_\nu b_\mu ) \delta^3$, which is needed
in proving the closure of the $ISOp(8|8)$ algebra among the generators, so the formulation depends on the Einstein-Hilbert term 
in a classical action. One of our motivations is to relax this situation and show that the choral symmetry does not 
depend on the expression of  the classical gravity but completely comes from the de Donder gauge condition for GCT
in the BRST formalism \cite{Kugo-Ojima}. For this purpose, without the knowledge of the classical Lagrangian we derive various 
equal-time (anti-)commutation relations (ETCRs) for the the Nakanishi-Lautrup field, the Faddeev-Popov (FP) ghost, 
and the FP antighost only on the basis of the de Donder gauge condition.

Another motivation behind the study at hand is to construct a quantum theory of the well-known Brans-Dicke gravity
\cite{Brans} by constructing its manifestly covariant BRST formalism  
since many of studies of the Brans-Dicke gravity have been limited to a classical analysis. As a concrete advantage 
of our quantum theory where the gravity as well as a scalar field are quantized, we will show that scale invariance is 
spontaneously broken and consequently ``dilaton'' is exactly massless thanks to the Nambu-Goldstone theorem 
even in quantum gravity.     
      
We close this section with an overview of this article. In Section 2, we discuss a quantum gravity for which the de
Donder gauge is adopted as a gauge condition for the general coordinate invariance. In Section 3, we calculate
the ETCRs for the Nakanishi-Lautrup auxiliary field, the FP ghost and the FP antighost based on the
quantum gravity made in Section 2. In Section 4, we calculate the ETCRs involving among the Nakanishi-Lautrup field
and its time derivative without using the information on a classical gravitational action, and then comment on choral
symmetry. In Section 5, a scale invariant scalar-tensor gravity is briefly reviewed. In Section 6, by selecting the scale 
invariant scalar-tensor gravity as a classical theory we calculate the ETCRs for the gravitational sector, and in Section 7 
we discuss spontaneous symmetry breakdown of the scale invariance and show that the dilaton is exactly massless 
owing to the Nambu-Goldstone theorem in quantum gravity. The final section is devoted to the conclusion. Two
appendices are put for technical details. In Appendix A, a derivation of the tensorlike ETCR is given, and in Appendix B
we have derived the scale current via the Noether theorem.

\section{Quantum gravity with de Donder gauge}

We wish to consider a manifestly covariant canonical formalism of general gravitational theories where
general coordinate invariance is fixed by the de Donder gauge condition. To take a more general theory into consideration, 
without specifying the concrete expression of the gravitational Lagrangian, we will start with the following classical 
Lagrangian:\footnote{We follow the notation and conventions of MTW textbook \cite{MTW}. Greek little letters $\mu, \nu, \cdots$ 
and Latin ones $i, j, \cdots$ are used for space-time and spatial indices, respectively; for instance, $\mu= 0, 1, 2, 3$ 
and $i = 1, 2, 3$. Furthermore, the Riemann curvature tensor and the Ricci tensor are respectively
defined by $R^\rho \, _{\sigma\mu\nu} = \partial_\mu \Gamma^\rho_{\sigma\nu} + \Gamma^\rho_{\lambda\mu} 
\Gamma^\lambda_{\sigma\nu} - ( \mu \leftrightarrow \nu)$ and $R_{\mu\nu} = R^\rho \, _{\mu\rho\nu}$.} 
\begin{eqnarray}
{\cal{L}}_c = {\cal{L}}_c ( g_{\mu\nu}, \phi ),
\label{Lc}  
\end{eqnarray}
which includes the metric tensor field $g_{\mu\nu}$ and a scalar field $\phi$ as dynamical fields, and is 
invariant under the general coordinate transformation (GCT). We assume that ${\cal{L}}_c$ does not 
involve more than first order derivatives of the metric and matter fields.

Let us fix the general coordinate symmetry by the de Donder gauge condition (or harmonic gauge condition):
\begin{eqnarray}
\partial_\mu\tilde g^{\mu\nu} = 0,
\label{Donder}  
\end{eqnarray}
where we have defined $\tilde g^{\mu\nu} \equiv \sqrt{-g} g^{\mu\nu} \equiv h g^{\mu\nu}$. Then, the BRST 
transformation is of form:
\begin{eqnarray}
\delta_B \bar c_\rho&=& i B_\rho, \quad \delta_B c^\rho= - c^\lambda\partial_\lambda c^\rho,
\quad \delta_B \phi = - c^\lambda \partial_\lambda \phi,
\nonumber\\
\delta_B g_{\mu\nu} &=& - ( \nabla_\mu c_\nu+ \nabla_\nu c_\mu)
\nonumber\\
&=& - ( c^\alpha\partial_\alpha g_{\mu\nu} + \partial_\mu c^\alpha g_{\alpha\nu} 
+ \partial_\nu c^\alpha g_{\mu\alpha} ),
\nonumber\\
\delta_B \tilde g^{\mu\nu} &=& h ( \nabla^\mu c^\nu+ \nabla^\nu c^\mu 
- g^{\mu\nu} \nabla_\rho c^\rho).
\label{BRST}  
\end{eqnarray}

Using this BRST transformation, the Lagrangian for the gauge-fixing condition and FP ghosts can be
constructed in a standard manner
\begin{eqnarray}
{\cal{L}}_{GF + FP} &=& \delta_B ( i \tilde g^{\mu\nu} \partial_\mu\bar c_\nu)
\nonumber\\
&=& - \tilde g^{\mu\nu} \partial_\mu B_\nu- i \partial_\mu\bar c_\nu\Bigl[ \tilde g^{\mu\rho} \partial_\rho c^\nu 
+ \tilde g^{\nu\rho} \partial_\rho c^\mu - \partial_\rho( \tilde g^{\mu\nu} c^\rho) \Bigr]. 
\label{GF+FP}  
\end{eqnarray}
To simplify this expression, let us introduce a new auxiliary field $b_\rho$ defined as
\begin{eqnarray}
b_\rho= B_\rho- i c^\lambda\partial_\lambda\bar c_\rho,
\label{b-field}  
\end{eqnarray}
and its BRST transformation reads
\begin{eqnarray}
\delta_B b_\rho= - c^\lambda\partial_\lambda b_\rho.
\label{b-BRST}  
\end{eqnarray}
Then, the Lagrangian (\ref{GF+FP}) can be cast to the form:
\begin{eqnarray}
{\cal{L}}_{GF + FP} = - \tilde g^{\mu\nu} \partial_\mu b_\nu- i \tilde g^{\mu\nu} \partial_\mu\bar c_\rho 
\partial_\nu c^\rho+ i \partial_\rho( \tilde g^{\mu\nu} \partial_\mu\bar c_\nu\cdot c^\rho). 
\label{GF+FP2}  
\end{eqnarray}

As a result, up to a total derivative, the gauge-fixed and BRST-invariant quantum Lagrangian is given by
\begin{eqnarray}
{\cal{L}}_q &=& {\cal{L}}_c - \tilde g^{\mu\nu} \partial_\mu b_\nu 
- i \tilde g^{\mu\nu} \partial_\mu\bar c_\rho\partial_\nu c^\rho
\nonumber\\
&\equiv& {\cal{L}}_c + {\cal{L}}_{GF} + {\cal{L}}_{FP},
\label{q-Lag}  
\end{eqnarray}
where we have defined
\begin{eqnarray}
{\cal{L}}_{GF} \equiv - \tilde g^{\mu\nu} \partial_\mu b_\nu, \qquad
{\cal{L}}_{FP} \equiv - i \tilde g^{\mu\nu} \partial_\mu\bar c_\rho\partial_\nu c^\rho.
\label{GF&FP}  
\end{eqnarray}
By performing the integration by parts once, let us rewrite the Lagrangian (\ref{q-Lag}) as
\begin{eqnarray}
{\cal{L}}_q &=& {\cal{L}}_c + \partial_\mu\tilde g^{\mu\nu} b_\nu 
- i \tilde g^{\mu\nu} \partial_\mu\bar c_\rho\partial_\nu c^\rho + \partial_\mu{\cal{V}}^\mu
\nonumber\\
&\equiv& {\cal{L}}_c + \bar {\cal{L}}_{GF} + {\cal{L}}_{FP} + \partial_\mu{\cal{V}}^\mu,
\label{Cov-Lag}  
\end{eqnarray}
where a surface term ${\cal{V}}^\mu$ and $\bar {\cal{L}}_{GF}$ are defined as
\begin{eqnarray}
{\cal{V}}^\mu \equiv - \tilde g^{\mu\nu} b_\nu, \quad
\bar {\cal{L}}_{GF} \equiv \partial_\mu\tilde g^{\mu\nu} b_\nu.
\label{surface}  
\end{eqnarray}

From this Lagrangian, we can obtain field equations by taking the variation with respect to $g_{\mu\nu}$,
$\phi$, $b_\nu$, $\bar c_\rho$ and $c^\rho$ in order:
\begin{eqnarray}
&{}& \frac{1}{\sqrt{-g}} \frac{\delta {\cal{L}}_c}{\delta g^{\mu\nu}} 
- \frac{1}{2} ( E_{\mu\nu} - \frac{1}{2} g_{\mu\nu} E ) = 0, \quad
\frac{\delta {\cal{L}}_c}{\delta \phi} = 0,
\nonumber\\
&{}& \partial_\mu\tilde g^{\mu\nu} = 0, \quad
g^{\mu\nu} \partial_\mu\partial_\nu c^\rho = 0, \quad
g^{\mu\nu} \partial_\mu\partial_\nu\bar c_\rho = 0,
\label{Field-eq}  
\end{eqnarray}
where we have defined 
\begin{eqnarray}
E_{\mu\nu}  &=& \partial_\mu b_\nu+ i \partial_\mu\bar c_\rho\partial_\nu c^\rho + ( \mu\leftrightarrow \nu),
\nonumber\\
E &=& g^{\mu\nu} E_{\mu\nu}.
\label{E}  
\end{eqnarray}

Next, in order to find the field equation for the $b_\rho$ field, let us take a covariant derivative of the first 
Einstein equation in Eq. (\ref{Field-eq}).  The result reads
\begin{eqnarray}
\nabla_\mu\left( E^{\mu\nu} - \frac{1}{2} g^{\mu\nu} E \right) = 0,
\label{Cov-E}  
\end{eqnarray}
where we have used the equation
\begin{eqnarray}
\nabla^\mu \frac{\delta {\cal{L}}_c}{\delta g^{\mu\nu}} = 0.
\label{Cov-g}  
\end{eqnarray}
This equation can be shown as follows: Using the GCT invariance of the classical action, we have
\begin{eqnarray}
0 &=& \delta_\varepsilon S_c \equiv \int d^4 x \delta_\varepsilon {\cal{L}}_c  
= \int d^4 x \Bigl( \frac{\delta {\cal{L}}_c}{\delta g^{\mu\nu}} \delta_\varepsilon g^{\mu\nu}
+ \frac{\delta {\cal{L}}_c}{\delta \phi} \delta_\varepsilon \phi \Bigr)
\nonumber\\
&=& \int d^4 x \frac{\delta {\cal{L}}_c}{\delta g^{\mu\nu}} \delta_\varepsilon g^{\mu\nu}
= \int d^4 x \frac{\delta {\cal{L}}_c}{\delta g^{\mu\nu}} (\nabla^\mu \varepsilon^\nu
+ \nabla^\nu \varepsilon^\mu )
\nonumber\\
&=& - 2 \int d^4 x \Bigl(\nabla^\mu \frac{\delta {\cal{L}}_c}{\delta g^{\mu\nu}} \Bigr) \varepsilon^\nu,
\label{Cov-S}  
\end{eqnarray}
where we have used the field equation for $\phi$.

Generally for a symmetric tensor $S^{\mu\nu}$, we have a formula:
\begin{eqnarray}
\nabla_\nu S^\nu{}_\mu= h^{-1}\partial_\nu(hS^\nu{}_\mu) +\frac12S_{\alpha\beta}\partial_\mu g^{\alpha\beta}.
\label{S-S}  
\end{eqnarray}
Using this, we can show an equality:
\begin{eqnarray}
\nabla_\nu E^\nu{}_\mu &=& 
h^{-1}\partial_\alpha\tilde g^{\alpha\nu}\cdot E_{\nu\mu}+ 
\partial^2 b_\mu +i\left(
\partial^2\bar c_\lambda\cdot \partial_\mu c^\lambda+\partial_\mu\bar c_\lambda\cdot \partial^2c^\lambda 
\right)
+\frac{1}{2} \partial_\mu E, 
\label{E-E}  
\end{eqnarray}
where $\partial^2 \equiv g^{\mu\nu}\partial_\mu\partial_\nu$.
Then Eq. (\ref{Cov-E}), together with the help of the other field equations in Eq. (\ref{Field-eq}), 
is seen to lead to the field equation for the $b_\rho$ field:
\begin{eqnarray}
g^{\mu\nu} \partial_\mu\partial_\nu b_\rho = 0.
\label{b-field-eq}  
\end{eqnarray}
In other words, the $b_\rho$ field, the ghost field $c^\rho$ and the antighost field $\bar c_\rho$ all satisfy 
the d'Alembert equation. Furthermore, it is of interest to see that the space-time coordinates $x^\lambda$ obey
the d'Alembert equation, $g^{\mu\nu} \partial_\mu\partial_\nu x^\lambda = 0$ as well.\footnote{Using the de Donder 
condition, these d'Alembert equations can be rewritten as $\partial_\mu ( \tilde g^{\mu\nu} \partial_\nu \Phi ) = 0$
where $\Phi \equiv \{ x^\lambda, b_\rho, c^\sigma, \bar c_\tau \}$.}

\section{Equal-time commutation relations}

In this section, after introducing the canonical commutation relations (CCRs), we will evaluate the equal-time 
commutation relations (ETCRs) among fundamental variables, in particular, the Nakanishi-Lautrup field $b_\mu$,
the FP ghost $c^\mu$ and the FP antighost $\bar c_\mu$ in detail. 
To simplify various expressions, we will obey the following abbreviations adopted in the textbook of Nakanishi
and Ojima \cite{N-O-text}:
\begin{eqnarray}
[ A, B^\prime ] &=& [ A(x), B(x^\prime) ] |_{x^0 = x^{\prime 0}},
\qquad \delta^3 = \delta(\vec{x} - \vec{x}^\prime), 
\nonumber\\
\tilde f &=& \frac{1}{\tilde g^{00}} = \frac{1}{\sqrt{-g} g^{00}} = \frac{1}{h g^{00}},
\label{abbreviation}  
\end{eqnarray}
where we assume that $\tilde g^{00}$ is invertible. 

Now let us set up the canonical (anti-)commutation relations: 
\begin{eqnarray}
[ g_{\mu\nu}, \pi_g^{\rho\lambda\prime} ] &=& i \frac{1}{2} ( \delta_\mu^\rho\delta_\nu^\lambda 
+ \delta_\mu^\lambda\delta_\nu^\rho) \delta^3,  \qquad [ \phi, \pi_\phi^\prime ] = + i \delta^3,
\nonumber\\
\{ c^\sigma, \pi_{c \lambda}^\prime,  \} &=& \{ \bar c_\lambda, \pi_{\bar c}^{\sigma\prime} \}
= + i \delta_\lambda^\sigma\delta^3,
\label{CCRs}  
\end{eqnarray}
where the other (anti-)commutation relations vanish.
Here the canonical variables are $g_{\mu\nu}, \phi, c^\rho, \bar c_\rho$ and the corresponding canonical
conjugate momenta are $\pi_g^{\mu\nu}, \pi_\phi, \pi_{c \rho}, \pi_{\bar c}^\rho$, respectively and 
the $b_\mu$ field is regarded as not a canonical variable but a conjugate momentum of $\tilde g^{0 \mu}$. 

Based on the Lagrangian (\ref{Cov-Lag}), the expressions for canonical conjugate momenta read
\begin{eqnarray}
\pi_g^{\mu\nu} &=& \frac{\partial{\cal{L}}}{\partial \dot g_{\mu\nu}}, 
\nonumber\\
\pi_\phi&=& \frac{\partial{\cal{L}}}{\partial \dot \phi},
\nonumber\\
\pi_{c \sigma} &=& \frac{\partial{\cal{L}}}{\partial \dot c^\sigma} = - i \tilde g^{\mu0} \partial_\mu\bar c_\sigma,
\nonumber\\
\pi_{\bar c}^\sigma&=& \frac{\partial{\cal{L}}}{\partial \dot {\bar c}_\sigma} = i \tilde g^{\mu0} \partial_\mu c^\sigma,
\label{CCM}  
\end{eqnarray}
where we have defined the time derivative such as $\dot g_{\mu\nu} \equiv \frac{\partial g_{\mu\nu}}{\partial t}
\equiv \partial_0 g_{\mu\nu}$, and differentiation of ghosts is taken from the right. 

From now on, we would like to evaluate various nontrivial equal-time commutation relations (ETCRs) in order.
Let us first work with the ETCR in Eq. (\ref{CCRs}):
\begin{eqnarray}
[ \pi_g^{\alpha0}, g_{\mu\nu}^\prime ] = - i \frac{1}{2} ( \delta_\mu^\alpha\delta_\nu^0 
+ \delta_\mu^0 \delta_\nu^\alpha) \delta^3.
\label{pi(a0)-g}  
\end{eqnarray}
The canonical conjugate momentum $\pi_g^{\alpha0}$ has a structure
\begin{eqnarray}
\pi_g^{\alpha0} = A^\alpha+ B^{\alpha\beta} \partial_\beta \phi+ C^{\alpha\beta} b_\beta,
\label{pi(a0)}  
\end{eqnarray}
where $A^\alpha, B^{\alpha\beta}$ and $C^{\alpha\beta} \equiv - \frac{1}{2} \tilde g^{00} g^{\alpha\beta}$ have 
no $\dot g_{\mu\nu}$, and $B^{\alpha\beta} \partial_\beta \phi$ does not have $\dot \phi$ 
since $\pi_g^{\alpha0}$ does not include the dynamics of the metric and the scalar fields. Then, we find 
that Eq. (\ref{pi(a0)-g}) produces
\begin{eqnarray}
[ g_{\mu\nu}, b_\rho^\prime ] = - i \tilde f ( \delta_\mu^0 g_{\rho\nu} + \delta_\nu^0 g_{\rho\mu} ) \delta^3.
\label{g-b}  
\end{eqnarray}
From this ETCR, we can easily derive ETCRs:
\begin{eqnarray}
&{}& [ g^{\mu\nu}, b_\rho^\prime ] = i \tilde f ( g^{\mu0} \delta_\rho^\nu+ g^{\nu0} \delta_\rho^\mu) \delta^3,
\nonumber\\
&{}& [ \tilde g^{\mu\nu}, b_\rho^\prime ] = i \tilde f ( \tilde g^{\mu0} \delta_\rho^\nu+ \tilde g^{\nu0} \delta_\rho^\mu 
- \tilde g^{\mu\nu} \delta_\rho^0 ) \delta^3.
\label{3-g-b}  
\end{eqnarray}
Here we have used the following fact; since a commutator works as a derivation, we can have formulae:
\begin{eqnarray}
&{}& [ g^{\mu\nu}, \Phi^\prime ] = - g^{\mu\alpha} g^{\nu\beta} [ g_{\alpha\beta}, \Phi^\prime ],
\nonumber\\
&{}& [ \tilde g^{\mu\nu}, \Phi^\prime ] = - \left( \tilde g^{\mu\alpha} g^{\nu\beta} - \frac{1}{2} \tilde g^{\mu\nu} 
g^{\alpha\beta} \right) [ g_{\alpha\beta}, \Phi^\prime ],
\label{Simple formulae}  
\end{eqnarray}
where $\Phi$ is a generic field.

As for the ETCRs involving ghosts, let us first consider the anti-ETCRs, $\{ \pi_{c \lambda}, c^{\sigma\prime} \}
= \{ \pi_{\bar c}^\sigma, \bar c_\lambda^\prime \} = i \delta_\lambda^\sigma\delta^3$. These ETCRs lead to
the same ETCR: 
\begin{eqnarray}
\{ \dot{\bar c}_\lambda, c^{\sigma\prime} \} = - \tilde f \delta_\lambda^\sigma\delta^3,
\label{gh-antigh}  
\end{eqnarray}
where we have used a useful identity for generic variables $\Phi$ and $\Psi$: 
\begin{eqnarray}
[ \Phi, \dot \Psi^\prime] = \partial_0 [ \Phi, \Psi^\prime] - [ \dot \Phi, \Psi^\prime],
\label{identity}  
\end{eqnarray}
which also holds for the anti-commutation relation. Next, it is easy to see that the ETCRs, 
$\{ \pi_g^{\alpha0}, c^{\sigma\prime} \} = \{ \pi_g^{\alpha0}, \bar c_\lambda^\prime \} = 0$ produce
\begin{eqnarray}
[ b_\rho, c^{\sigma\prime} ] = [ b_\rho, \bar c_\lambda^\prime ] = 0.
\label{b-ghs}  
\end{eqnarray}
Finally, the ETCRs, $[ \pi_{c \lambda}, \pi_g^{\alpha0 \prime} ] = [ \pi_{\bar c}^\sigma, \pi_g^{\alpha0 \prime} ]
= 0$ give us  
\begin{eqnarray}
[ \dot{\bar c}_\lambda, b_\rho^\prime ] = - i \tilde f \partial_\rho\bar c_\lambda\delta^3, \qquad
[ \dot c^\sigma, b_\rho^\prime ] = - i \tilde f \partial_\rho c^\sigma\delta^3.
\label{c-b}  
\end{eqnarray}

In this article, we make use of the following ETCR:
\begin{eqnarray}
[ \dot g_{\mu\nu}, b_\rho^\prime ] &=& - i \Bigl\{ \tilde f ( \partial_\rho g_{\mu\nu} + \delta_\mu^0 \dot g_{\rho\nu} 
+ \delta_\nu^0 \dot g_{\rho\mu} ) \delta^3 
\nonumber\\
&+& [ ( \delta_\mu^k - 2 \delta_\mu^0 \tilde f \tilde g^{0 k} ) g_{\rho\nu}
+ (\mu\leftrightarrow \nu) ] \partial_k ( \tilde f \delta^3 ) \Bigr\}. 
\label{dot g-b}  
\end{eqnarray}
This ETCR can be in general shown to hold when the system has the translational invariance and the general coordinate
transformation is fixed by the de Donder gauge as follows:
The translational invariance requires the validity of the following equation for a generic field $\Phi(x)$:
\begin{eqnarray}
[ \Phi(x), P_\rho ] = i \partial_\rho \Phi(x),
\label{Translation}  
\end{eqnarray}
where $P_\rho$ is the generator of the translation which is now given by\footnote{This translation generator belongs
to the generators of choral symmetry as will be seen in Eq. (\ref{Choral-gen}). }
\begin{eqnarray}
P_\rho = \int d^3 x \, \tilde g^{0\lambda} \partial_\lambda b_\rho.
\label{P}  
\end{eqnarray}

Now let us consider the specific case $\Phi(x) = g_{\mu\nu}(x)$:
\begin{eqnarray}
[ g_{\mu\nu}(x), P_\rho ] = [ g_{\mu\nu}(x), \int d^3 x' \, \tilde g^{0\lambda\prime} 
\partial_\lambda b_\rho^\prime ] = i \partial_\rho g_{\mu\nu}(x).
\label{g-translation}  
\end{eqnarray}  
Taking $x^0 = x^{\prime0}$ and using $[ g_{\mu\nu}, \tilde g^{0\lambda \prime} ] = 0$, we have
\begin{eqnarray}
\int d^3 x' \, \tilde g^{0\lambda}(x^\prime)  [ g_{\mu\nu}, \partial_\lambda b_\rho^\prime ] 
= i \partial_\rho g_{\mu\nu}(x).
\label{g-translation2}  
\end{eqnarray}  
Using the de Donder gauge condition (\ref{Donder}) and Eq. (\ref{g-b}), this equation can be rewritten as
\begin{eqnarray}
\int d^3 x' \, \tilde g^{00}(x^\prime)  [ g_{\mu\nu}, \dot b_\rho^\prime ] 
= i \left[ \partial_\rho g_{\mu\nu} - \frac{1}{\tilde f} \partial_0 \tilde f ( \delta_\mu^0 g_{\rho\nu}
+ \delta_\nu^0 g_{\rho\mu} ) \right],
\label{g-translation3}  
\end{eqnarray}  
which is easily solved for $[ g_{\mu\nu}, \dot b_\rho^\prime ]$ to be
\begin{eqnarray}
[ g_{\mu\nu}, \dot b_\rho^\prime ] 
= i \left[ \tilde f \partial_\rho g_{\mu\nu} - \partial_0 \tilde f ( \delta_\mu^0 g_{\rho\nu}
+ \delta_\nu^0 g_{\rho\mu} ) \right] \delta^3 + F_{(\mu\nu)\rho} \,^k \partial_k (\tilde f \delta^3),
\label{g-dot-b-sol}  
\end{eqnarray}  
where $F_{(\mu\nu)\rho} \,^k$ is an arbitrary function which is symmetric under the exchange of $\mu
\leftrightarrow \nu$.

Next, to fix the function $F_{(\mu\nu)\rho} \,^k$, let us take account of the consistency with the de Donder 
gauge condition (\ref{Donder}):
\begin{eqnarray}
[ \partial_\mu \tilde g^{\mu\nu}, b_\rho^\prime ] = 0.
\label{Consist-Donder}  
\end{eqnarray}  
After some calculations, Eq. (\ref{Consist-Donder}) leads to an equation for $F_{(\mu\nu)\rho} \,^k$:
\begin{eqnarray}
\left( \tilde g^{0\alpha} g^{\nu\beta} - \frac{1}{2} \tilde g^{0\nu} g^{\alpha\beta} \right) F_{(\alpha\beta)\rho} \,^k
= - i ( \tilde g^{0k} \delta_\rho^\nu + \tilde g^{0\nu} \delta_\rho^k - \tilde g^{k\nu} \delta_\rho^0 ).
\label{Consist-Donder2}  
\end{eqnarray}  
This equation has the unique solution given by 
\begin{eqnarray}
F_{(\mu\nu)\rho} \,^k
= i [ ( \delta_\mu^k - 2 \delta_\mu^0 \tilde f \tilde g^{0k} ) g_{\rho\nu} + (\mu \leftrightarrow \nu) ].
\label{F-sol}  
\end{eqnarray}  
We can therefore obtain
\begin{eqnarray}
[ g_{\mu\nu}, \dot b_\rho^\prime ] 
&=& i \Bigl\{ [ \tilde f \partial_\rho g_{\mu\nu} - \partial_0 \tilde f ( \delta_\mu^0 g_{\rho\nu}
+ \delta_\nu^0 g_{\rho\mu} ) ] \delta^3 
\nonumber\\
&+& [ ( \delta_\mu^k - 2 \delta_\mu^0 \tilde f \tilde g^{0k} ) g_{\rho\nu} + (\mu \leftrightarrow \nu) ]
\partial_k (\tilde f \delta^3) \Bigr\}.
\label{g-dot-b-final}  
\end{eqnarray}  
Finally, using Eqs. (\ref{g-b}) and (\ref{identity}), we can arrive at the desired equation (\ref{dot g-b}). 
It is of interest that Eq. (\ref{dot g-b}) can be derived from only the translational invariance and 
the de Donder gauge condition without reference to the classical Lagrangian ${\cal{L}}_c$ which knows 
information of the dynamics of the gravitational field $g_{\mu\nu}$ and the scalar field $\phi$.

Then, using Eq. (\ref{dot g-b}) together with Eqs. (\ref{g-b}) and (\ref{3-g-b}), we can easily show that
\begin{eqnarray}
&{}& [ \Gamma^\rho_{\mu\nu}, b_\lambda^\prime ] = i \tilde f ( \delta_\lambda^\rho \Gamma_{\mu\nu}^0 
- \delta_\mu^0 \Gamma_{\lambda\nu}^\rho - \delta_\nu^0 \Gamma_{\mu\lambda}^\rho ) \delta^3
\nonumber\\
&{}& + i \delta_\lambda^\rho ( 2 \delta^0_\mu \delta^0_\nu \tilde f \tilde g^{0k} - \delta^0_\mu \delta^k_\nu
- \delta^0_\nu \delta^k_\mu ) \partial_k ( \tilde f \delta^3 ).
\label{Gamma-b}  
\end{eqnarray}
Incidentally, this ETCR is also needed in deriving the ETCR $[ b_\mu, \dot b_\nu^\prime ]$
in the next section and the tensorlike ETCR 
\begin{eqnarray}
[ R_{\mu\nu}, b_\rho^\prime ] = - i \tilde f ( \delta_\mu^0 R_{\rho\nu}
+  \delta_\nu^0 R_{\rho\mu} ) \delta^3,
\label{R-b}  
\end{eqnarray}
which will be derived in the Appendix A.

Finally, for later conveniece we derive the ETCRs, $[ \phi, b_\rho^\prime ]$, $[ \dot \phi, b_\rho^\prime ]$
and $[ \dot \phi, \dot b_\rho^\prime ]$.
To do so, we begin by considering $[ \phi, \bar c_\rho^\prime ] = 0$ and take its BRST transformation
as follows:\footnote{We define the BRST transformation as $\delta_B \Phi \equiv [ i Q_B, \Phi \}$ where 
$\Phi$ is a generic field and $[ \hspace{2mm}, \hspace{2mm} \}$ denotes the graded bracket.}
\begin{eqnarray}
0 &=& \{ i Q_B, [ \phi, \bar c_\rho^\prime ] \} 
= \{ [ i Q_B, \phi ], \bar c_\rho^\prime \} + [ \phi, \{ i Q_B, \bar c_\rho^\prime \} ]
\nonumber\\
&=& \{ - c^\lambda \partial_\lambda \phi, \bar c_\rho^\prime \} + [ \phi, i B_\rho^\prime ] 
= [ \phi, i ( b_\rho^\prime + i c^{\lambda\prime} \partial_\lambda \bar c_\rho^\prime ) ]
\nonumber\\
&=& i [ \phi, b_\rho^\prime ],
\label{Phi-b1}  
\end{eqnarray}
where together with $[ \dot \phi, \bar c_\rho^\prime ] = 0$, Eqs. (\ref{BRST}) and
(\ref{b-field}) have been used.
Thus, we have shown 
\begin{eqnarray}
[ \phi, b_\rho^\prime ] = 0.
\label{Phi-b2}  
\end{eqnarray}
In a perfectly similar way, we can calculate $[ \dot \phi, b_\rho^\prime ]$ by starting with 
$[ \dot \phi, \bar c_\rho^\prime ] = 0$, which holds in the de Donder gauge, as follows:
\begin{eqnarray}
0 &=& \{ i Q_B, [ \dot \phi, \bar c_\rho^\prime ] \} 
= \{ [ i Q_B, \dot \phi ], \bar c_\rho^\prime \} + [ \dot \phi, \{ i Q_B, \bar c_\rho^\prime \} ]
\nonumber\\
&=& \{ - \partial_0 (c^\lambda \partial_\lambda \phi), \bar c_\rho^\prime \} 
+ [ \dot \phi, i ( b_\rho^\prime + i c^{\lambda\prime} \partial_\lambda \bar c_\rho^\prime ) ] 
\nonumber\\
&=& - \{ \dot c^\lambda, \bar c_\rho^\prime \} \partial_\lambda \phi 
+ i [ \dot \phi, b_\rho^\prime ],
\label{dot-phi-b}  
\end{eqnarray}  
from which, using Eq. (\ref{gh-antigh}) we can obtain 
\begin{eqnarray}
[ \dot \phi, b_\rho^\prime ] = - i \tilde f \partial_\rho \phi \delta^3.
\label{dot-phi-b2}  
\end{eqnarray}

The calculation of $[ \dot \phi, \dot b_\rho^\prime ]$ proceeds as follows: First, we utilize
the formula (cf. Eq. (\ref{identity}))
\begin{eqnarray}
[ \dot \phi, \dot b_\rho^\prime ] = \partial_0 ( [ \phi, \dot b_\rho^\prime ] )
- [ \phi, \ddot b_\rho^\prime ].
\label{dot-phi-dot-b1}  
\end{eqnarray}
Since the $b$ field obeys the d'Alembert equation as in Eq. (\ref{b-field-eq}), $\ddot b_\rho^\prime$
can be described in terms of $\dot b_\rho$ and $b_\rho$ like 
\begin{eqnarray}
\ddot b_\rho = - \tilde f ( 2 \tilde g^{0k} \partial_i \dot b_\rho + \tilde g^{kl} 
\partial_k \partial_l b_\rho ).
\label{ddot-b}  
\end{eqnarray}
Then, we are ready to evaluate 
\begin{eqnarray}
[ \phi, \ddot b_\rho^\prime ] &=& - 2 \tilde f^\prime \tilde g^{0k\prime} \partial_k^\prime
[ \phi, \dot b_\rho^\prime ] = - 2 i \tilde f^\prime \tilde g^{0k\prime} \partial_k^\prime
( \tilde f \partial_\rho \phi \delta^3 )
\nonumber\\
&=& 2 i \tilde f \partial_\rho \phi \left[ \tilde f^{-1} \partial_0 \tilde f \delta^3 
+ \tilde g^{0k} \partial_k ( \tilde f \delta^3 ) \right].
\label{dot-phi-dot-b2}  
\end{eqnarray}
Thus, we reach a result
\begin{eqnarray}
[ \dot \phi, \dot b_\rho^\prime ] = i \tilde f ( \partial_\rho \dot \phi - \tilde f^{-1} \partial_0 \tilde f 
\partial_\rho \phi ) \delta^3 
- 2 i \tilde f \partial_\rho \phi \tilde g^{0k} \partial_k ( \tilde f \delta^3 ).
\label{dot-phi-dot-b3}  
\end{eqnarray}
Note that Eqs. (\ref{Phi-b2}), (\ref{dot-phi-b2}) and (\ref{dot-phi-dot-b3}) hold as well when 
we replace $\phi$ by $\varphi$.

\section{Derivation of ETCRs involving $b_\mu$ field and choral symmetry}

We are now in a position to address the novel part of our formulation and discuss a huge
residual symmetry called ``choral symmetry'' which emerges in adopting the de Donder gauge
for the general coordinate invariance.

In order to derive the commutation relations among the $b_\mu$ field in terms of 
the BRST transformation, let us start with the latter equation in Eq. (\ref{b-ghs}) and take 
the BRST transformation:
\begin{eqnarray}
0 &=& \{ i Q_B, [ b_\mu, \bar c_\nu^\prime ] \} 
\nonumber\\
&=& \{ [ i Q_B, b_\mu], \bar c_\nu^\prime \} + [ b_\mu, \{ i Q_B, \bar c_\nu^\prime \} ]
\nonumber\\
&=& - \{ c^\rho\partial_\rho b_\mu, \bar c_\nu^\prime \} + i [ b_\mu, b_\nu^\prime ] 
- [ b_\mu, c^{\rho\prime} \partial_\rho\bar c_\nu^\prime ], 
\label{b-b-1}  
\end{eqnarray}
where Eqs. (\ref{BRST}), (\ref{b-field}) and (\ref{b-BRST}) have been used.   
Using Eq. (\ref{c-b}), the first and third terms precisely cancel so we can obtain
\begin{eqnarray}
[ b_\mu, b_\nu^\prime ] = 0.
\label{b-b}  
\end{eqnarray}
  
Next, let us turn our attention to the derivation of the ETCR:
\begin{eqnarray}
[ b_\mu, \dot b_\nu^\prime ] = i \tilde f ( \partial_\mu b_\nu+ \partial_\nu b_\mu) \delta^3.
\label{b-dot-b}  
\end{eqnarray}
To do that, let us start with the ETCR
\begin{eqnarray}
[ \pi_{c \mu}, b_\nu^\prime ] = 0,
\label{pi-b}  
\end{eqnarray}
which can be easily shown. Taking its BRST transformation leads to an equation:
\begin{eqnarray}
[ \{ i Q_B, \pi_{c \mu} \}, b_\nu^\prime ] - \{ \pi_{c \mu}, [ i Q_B, b_\nu^\prime ] \} = 0.
\label{pi-b-BRST}  
\end{eqnarray}
The first term on the LHS can be calculated to be
\begin{eqnarray}
&{}& [ \{ i Q_B, \pi_{c \mu} \}, b_\nu^\prime ] = - i [ ( \tilde g^{\rho\sigma} \nabla_\sigma c^0 
+ \tilde g^{0 \sigma} \nabla_\sigma c^\rho - \tilde g^{\rho 0} \nabla_\lambda c^\lambda ) 
\partial_\rho \bar c_\mu, b_\nu^\prime ]
\nonumber\\
&{}& + i \partial_\nu b_\mu \cdot \delta^3 - \tilde g^{00} [ b_\mu, \dot b_\nu^\prime ] 
+ i [ \tilde g^{\rho 0} \partial_\rho ( c^\lambda \partial_\lambda \bar c_\mu ), b_\nu^\prime ], 
\label{pi-b-BRST1}  
\end{eqnarray}
where Eqs. (\ref{BRST}), (\ref{b-field}), (\ref{CCM}), (\ref{3-g-b}) and (\ref{b-b}) have been used.
The second term on the LHS in Eq. (\ref{pi-b-BRST}) can be also calculated to be
\begin{eqnarray}
\{ \pi_{c \mu}, [ i Q_B, b_\nu^\prime ] \} = - i \partial_\mu b_\nu \cdot \delta^3 + c^{\lambda \prime} 
[ \pi_{c \mu}, \partial_\lambda b_\nu^\prime ].
\label{pi-b-BRST2}  
\end{eqnarray}
Then, Eq. (\ref{pi-b-BRST}) together with Eqs. (\ref{pi-b-BRST1}) and (\ref{pi-b-BRST2}) gives us an equation:
\begin{eqnarray}
[ b_\mu, \dot b_\nu^\prime ] &=& i \tilde f ( \partial_\mu b_\nu + \partial_\nu b_\mu) \delta^3
- i \tilde f [ ( \tilde g^{\rho\sigma} \nabla_\sigma c^0 + \tilde g^{0 \sigma} \nabla_\sigma c^\rho
- \tilde g^{\rho 0} \nabla_\lambda c^\lambda ) \partial_\rho \bar c_\mu, b_\nu^\prime ]
\nonumber\\
&+& i \tilde f [ \tilde g^{\rho 0} \partial_\rho ( c^\lambda \partial_\lambda \bar c_\mu ), b_\nu^\prime ]
- \tilde f c^{\lambda \prime} [ \pi_{c \mu}, \partial_\lambda b_\nu^\prime ]
\nonumber\\
&\equiv& i \tilde f ( \partial_\mu b_\nu + \partial_\nu b_\mu) \delta^3 + K_{\mu\nu}.
\label{pre-b-dot-b}  
\end{eqnarray}
After some calculations, we can prove $K_{\mu\nu} = 0$, which implies that Eq. (\ref{b-dot-b}) is certainly valid.
Note that in proving $K_{\mu\nu} = 0$, it is necessary to make use of Eq. (\ref{Gamma-b}) and the field equation for
the antighost $\bar c_\mu$, i.e., $\tilde g^{\mu\nu} \partial_\mu \partial_\nu \bar c_\rho = 0$.
In particular, we must use the following ETCRs:
\begin{eqnarray}
&{}& [ \ddot{\bar c}_\mu , b_\nu^\prime ] = - 2 i \tilde f [  \partial_\nu \dot{\bar c}_\mu \delta^3 
- \tilde g^{0k} \partial_\nu \bar c_\mu \partial_k ( \tilde f \delta^3 ) ],
\nonumber\\
&{}& [ \nabla_\sigma c^\rho, b_\nu^\prime ] = i \tilde f ( - \delta_\sigma^0 \nabla_\nu c^\rho 
+ \delta_\nu^\rho \Gamma_{\sigma\lambda}^0 c^\lambda - \Gamma_{\sigma\nu}^\rho c^0 ) \delta^3
\nonumber\\
&{}& + i \delta_\nu^\rho ( 2 \delta_\sigma^0 c^0 \tilde f \tilde g^{0k} - \delta_\sigma^0 c^k - \delta_\sigma^k c^0 )
\partial_k ( \tilde f \delta^3 ).
\label{nabla-c-b}  
\end{eqnarray}  

We end this section with the argument of choral symmetry, which is a huge residual symmetry $IOSp(8|8)$ 
involving the BRST symmetry, the rigid translation and $GL(4)$ symmetry etc.  
Via the Noether theorem, the $IOSp(8|8)$ generators can be constructed out of ${\cal{L}}_{GF} + {\cal{L}}_{FP}$ 
in Eq. (\ref{GF&FP}) as \cite{N-O-text}
\begin{eqnarray}
M^{MN} &\equiv& \int d^3 x \, \tilde g^{0\nu} ( X^M \overleftrightarrow{\partial}_\nu X^N )
\equiv \int d^3 x \, \tilde g^{0\nu} ( X^M \partial_\nu X^N - \partial_\nu X^M \cdot X^N ),
\nonumber\\
P^M &\equiv& \int d^3 x \, \tilde g^{0\nu} ( 1 \overleftrightarrow{\partial}_\nu X^N )
\equiv \int d^3 x \, \tilde g^{0\nu} \partial_\nu X^M,
\label{Choral-gen}  
\end{eqnarray}
where $X^M \equiv \{ x^\mu, b_\mu, c^\mu, \bar c_\mu \}$.
The $IOSp(8|8)$ algebra takes the graded form:
\begin{eqnarray}
&{}& [ i M^{MN}, M^{RS} \} = - M^{MS} \tilde \eta^{NR} + (-)^{|R||S|} M^{MR} \tilde \eta^{NS}
- (-)^{|M||N|} ( M \leftrightarrow N ),
\nonumber\\
&{}& [ i M^{MN}, P^R \} = - P^M \tilde \eta^{NR} + (-)^{|M||N|} P^N \tilde \eta^{MR},  \qquad
[ P^M, P^N \} = 0,
\label{Choral-algebra}  
\end{eqnarray}
where $\tilde \eta^{MN}$ is a $16 \times 16$ $IOSp(8|8)$ metric \cite{Kugo}.

Since the generators $M^{MN}$ and $P^M$ could have one time derivative, calculating the algebra
requires us to use the ETCRs including two time derivatives such as
\begin{eqnarray}
&{}& [ \dot b_\mu, \dot b_\nu^\prime ] = i \tilde f \left[ \partial_\mu \dot b_\nu + \partial_\nu \dot b_\mu
- 2 \tilde f \tilde g^{0\rho} \partial_\rho \partial_\mu b_\nu - \frac{\partial_0 \tilde f}{\tilde f}
( \partial_\mu b_\nu + \partial_\nu b_\mu ) \right] \delta^3,
\nonumber\\
&{}& - 2 i \tilde f \tilde g^{0k} ( \partial_\mu b_\nu + \partial_\nu b_\mu ) \partial_k ( \tilde f \delta^3)
\nonumber\\
&{}& [ \dot c^\mu, \dot b_\nu^\prime ] = i \tilde f \left[ \left( \partial_\nu \dot c^\mu 
- \frac{\partial_0 \tilde f}{\tilde f} \partial_\nu c^\mu \right) \delta^3 
- 2 \tilde g^{0k} \partial_\nu c^\mu \partial_k ( \tilde f \delta^3) \right], 
\nonumber\\
&{}& [ \dot{\bar c}_\mu, \dot b_\nu^\prime ] = i \tilde f \left[ \left( \partial_\nu \dot{\bar c}_\mu 
- \frac{\partial_0 \tilde f}{\tilde f} \partial_\nu \bar c_\mu \right) \delta^3 
- 2 \tilde g^{0k} \partial_\nu \bar c_\mu \partial_k ( \tilde f \delta^3) \right].
\label{dot-dot-CR}  
\end{eqnarray}

These ETCRs can be all derived from the field equations and the ETCRs obtained so far, for instance, Eqs. (\ref{Field-eq}), 
(\ref{b-field-eq}), (\ref{identity}), (\ref{b-ghs}) and (\ref{c-b}) without specifying the expression of a classical 
gravitational Lagrangian. This situation should be contrasted  with the previous formulation \cite{N-O-text}
where Eq. (\ref{b-dot-b}) has been derived by using the Einstein-Hilbert Lagrangian. Thus, it is said that 
the choral symmetry uniquely characterizes the expression of ${\cal{L}}_{GF} + {\cal{L}}_{FP}$ \cite{N-O-text}.
By contrast, we can mention that the origin of the choral symmetry purely lies in the de Donder gauge 
and the corresponding FP ghost Lagrangian irrespective of a specific choice of a classical gravitational theory
as long as there exists the general coordinate invariance in the classical theory.

\section{A scale invariant scalar-tensor gravity}

In this section, as a classical Lagrangian, we will take the simplest scalar-tensor gravity \cite{Fujii} whose Lagrangian 
is given by
\begin{eqnarray}
{\cal{L}}_c = \sqrt{-g} \left( \frac{1}{2} \xi \phi^2 R - \frac{1}{2} g^{\mu\nu} \partial_\mu \phi \partial_\nu \phi \right),
\label{ST-gravity}  
\end{eqnarray}
where $\xi$ is a constant called the non-minimal coupling constant, $\phi$ a real scalar field with a normal kinetic term 
(i.e., not a ghost), and $R$ the scalar curvature.
In addition to the general coordinate transformation (GCT) and a global scale transformation with $\Omega = \textrm{constant}$, 
this Lagrangian is also invariant under the restricted Weyl transformation \cite{Edery1}-\cite{Edery3}, 
\cite{Oda-R}-\cite{Oda-RWS}: 
\begin{eqnarray}
g_{\mu\nu} \rightarrow g^\prime_{\mu\nu} = \Omega^2 (x) g_{\mu\nu}, \qquad 
\phi \rightarrow \phi^\prime = \Omega^{-1}(x) \phi, 
\label{Res-Weyl}  
\end{eqnarray}
where the gauge transformation parameter $\Omega(x)$ obeys a constraint $\Box \Omega = 0$.
In order to prove the invariance, we need to use the following transformation of the scalar curvature under (\ref{Res-Weyl}):
\begin{eqnarray}
R \rightarrow R^\prime = \Omega^{-2} ( R - 6 \Omega^{-1} \Box \Omega ). 
\label{Weyl-R}  
\end{eqnarray}

For the sake of simplicity, in what follows we will put 
\begin{eqnarray}
\varphi \equiv \frac{1}{2} \xi \phi^2. 
\label{varphi}  
\end{eqnarray}
It is worth recalling that we can rewrite (\ref{ST-gravity}) as the Lagrangian of the well-known Brans-Dicke 
theory \cite{Brans}:
\begin{eqnarray}
{\cal{L}}_c = \sqrt{-g} \left( \varphi R - \omega \frac{1}{\varphi} g^{\mu\nu} \partial_\mu \varphi 
\partial_\nu \varphi \right),
\label{BD-Lag}  
\end{eqnarray}
where $\omega \equiv \frac{1}{4 \xi}$ is known as the Brans-Dicke parameter. Thus, our classical 
Lagrangian (\ref{ST-gravity}) is at least classically equivalent to that of Brans-Dicke theory.  
 
Taking the de Donder condition as a gauge-fixing condition for GCT, the gauge-fixed and 
BRST-invariant quantum Lagrangian is given by
\begin{eqnarray}
{\cal{L}}_q = \sqrt{-g} \left( \varphi R - \frac{1}{2} g^{\mu\nu} \partial_\mu \phi \partial_\nu \phi \right)
 - \tilde g^{\mu\nu} \partial_\mu b_\nu 
- i \tilde g^{\mu\nu} \partial_\mu \bar c_\rho \partial_\nu c^\rho.
\label{ST-q-Lag}  
\end{eqnarray}
From this Lagrangian, we can obtain field equations by taking the variation with respect to $g_{\mu\nu}$,
$\phi$, $b_\nu$, $\bar c_\rho$ and $c^\rho$ in order:
\begin{eqnarray}
&{}& \varphi G_{\mu\nu} - ( \nabla_\mu \nabla_\nu - g_{\mu\nu} \Box ) \varphi - \frac{1}{2} T_{\mu\nu} 
- \frac{1}{2} ( E_{\mu\nu} - \frac{1}{2} g_{\mu\nu} E ) = 0, 
\nonumber\\
&{}& \xi \phi R + \Box \phi = 0, \qquad
\partial_\mu\tilde g^{\mu\nu} = 0, 
\nonumber\\
&{}& g^{\mu\nu} \partial_\mu\partial_\nu c^\rho = 0, \qquad
g^{\mu\nu} \partial_\mu\partial_\nu\bar c_\rho = 0,
\label{q-Field-eq}  
\end{eqnarray}
where we have defined 
\begin{eqnarray}
T_{\mu\nu}  = \partial_\mu \phi \partial_\nu \phi - \frac{1}{2} g_{\mu\nu} (\partial_\rho \phi)^2.
\label{Def-T}  
\end{eqnarray}
Of course, even in this case the $b_\rho$ field satisfies the d'Alembert equation (\ref{b-field-eq}).

\section{Equal-time commutation relations in gravitational sector}

Since we have introduced the classical Lagrangian (\ref{ST-gravity}) in a theory at hand, 
we are now ready to evaluate the equal-time commutation relations (ETCRs) involving the metric 
tensor and the scalar fields in the gravitational sector. 

For later convenience, here let us take account of the de Donder gauge condition (\ref{Donder}), 
from which we have identities:
\begin{eqnarray}
g^{\mu\nu} \Gamma^\lambda_{\mu\nu} = 0, \qquad  g^{\lambda\mu} \partial_\lambda g_{\mu\nu} 
= \Gamma^\lambda_{\lambda\nu}.
\label{Donder-iden}  
\end{eqnarray}
Moreover, since the equation $g^{\mu\nu} \Gamma^\lambda_{\mu\nu} = 0$ reads
\begin{eqnarray}
( 2 g^{\lambda\mu} g^{\nu \rho} - g^{\mu\nu} g^{\lambda \rho} ) \partial_\rho g_{\mu\nu} = 0, 
\label{Donder-iden2}  
\end{eqnarray}
it is possible to express the time derivative of the metric field in terms of its spacial one as
\begin{eqnarray}
{\cal{D}}^{\lambda\mu\nu} \dot g_{\mu\nu} = ( 2 g^{\lambda\mu} g^{\nu k} - g^{\mu\nu} g^{\lambda k} )
\partial_k g_{\mu\nu}, 
\label{D-eq}  
\end{eqnarray}
where the operator ${\cal{D}}^{\lambda\mu\nu}$ is defined by
\begin{eqnarray}
{\cal{D}}^{\lambda\mu\nu} = g^{0 \lambda} g^{\mu\nu} - 2 g^{\lambda\mu} g^{0 \nu}.
\label{D-op}  
\end{eqnarray}

To remove second order derivatives of the metric involved in $R$, we perform the integration by parts once and
rewrite the Lagrangian (\ref{ST-q-Lag}) as\footnote{See Appendix B.}
\begin{eqnarray}
{\cal{L}} &=& - \varphi \tilde g^{\mu\nu} ( \Gamma^\sigma_{\mu\nu} \Gamma^\alpha_{\sigma\alpha}  
-  \Gamma^\sigma_{\mu\alpha} \Gamma^\alpha_{\sigma\nu} ) - \partial_\mu \varphi ( \tilde g^{\alpha\beta} 
\Gamma^\mu_{\alpha\beta}  - \tilde g^{\mu\nu} \Gamma^\alpha_{\nu\alpha} ) 
\nonumber\\
&-& \frac{1}{2} \tilde g^{\mu\nu} \partial_\mu \phi \partial_\nu \phi + \partial_\mu \tilde g^{\mu\nu} b_\nu 
- i \tilde g^{\mu\nu} \partial_\mu \bar c_\rho \partial_\nu c^\rho + \partial_\mu {\cal{V}}^\mu,
\label{ST-q-Lag2}  
\end{eqnarray}
where a surface term ${\cal{V}}^\mu$ is defined as
\begin{eqnarray}
{\cal{V}}^\mu =  \varphi ( \tilde g^{\alpha\beta} \Gamma^\mu_{\alpha\beta} - \tilde g^{\mu\nu} 
\Gamma^\alpha_{\nu\alpha} ) - \tilde g^{\mu\nu} b_\nu.
\label{surface2}  
\end{eqnarray}

From the Lagrangian (\ref{ST-q-Lag2}), the concrete expressions for canonical conjugate momenta for the metric tensor
and the scalar fields read
\begin{eqnarray}
\pi_g^{\mu\nu} &=& \frac{\partial {\cal{L}}}{\partial \dot g_{\mu\nu}} 
\nonumber\\
&=& - \frac{1}{2} \sqrt{-g} \, \varphi \Bigl[ - g^{0 \lambda} g^{\mu\nu} g^{\sigma\tau} - g^{0 \tau} g^{\mu\lambda} g^{\nu\sigma}
- g^{0 \sigma} g^{\mu\tau} g^{\nu\lambda} + g^{0 \lambda} g^{\mu\tau} g^{\nu\sigma} 
\nonumber\\
&+& g^{0 \tau} g^{\mu\nu} g^{\lambda\sigma}
+ \frac{1}{2} ( g^{0 \mu} g^{\nu\lambda} + g^{0 \nu} g^{\mu\lambda} ) g^{\sigma\tau} \Biggr] \partial_\lambda g_{\sigma\tau}
\nonumber\\
&-& \sqrt{-g} \Biggl[ \frac{1}{2} ( g^{0 \mu} g^{\rho\nu} + g^{0 \nu} g^{\rho\mu} ) - g^{\mu\nu} g^{\rho 0} \Bigr] 
\partial_\rho \varphi
\nonumber\\
&-& \frac{1}{2} \sqrt{-g} ( g^{0 \mu} g^{\nu\rho} + g^{0 \nu} g^{\mu\rho} - g^{0 \rho} g^{\mu\nu} )  b_\rho,
\nonumber\\
\pi_\phi &=& \frac{\partial {\cal{L}}}{\partial \dot \phi} = - \tilde g^{0 \mu} \partial_\mu \phi
+ \xi \phi ( - \tilde g^{\alpha\beta} \Gamma^0_{\alpha\beta} + \tilde g^{0 \nu} \Gamma^\alpha_{\nu\alpha} ).
\label{Grav-CCM}  
\end{eqnarray}

From now on, we would like to evaluate several nontrivial equal-time commutation relations (ETCRs) relevant to 
the gravitational Lagrangian (\ref{ST-q-Lag2}). For this purpose, let us first write down some equations to be solved in order.
Since $\pi_\phi$ in Eq. (\ref{Grav-CCM}) is rewritten as
\begin{eqnarray}
\pi_\phi = - \tilde g^{0 0} \dot \phi - \tilde g^{0 k} \partial_k \phi 
+ \xi \phi \left[ ( \tilde g^{0 0} g^{\rho\sigma} - \tilde g^{0 \rho} g^{0 \sigma} ) \dot g_{\rho\sigma}
+ ( \tilde g^{0 k} g^{\rho\sigma} - \tilde g^{0 \rho} g^{k \sigma} ) \partial_k g_{\rho\sigma} \right],
\label{pi-phi}  
\end{eqnarray}
$[ \pi_\phi, \phi^\prime ] = - i \delta^3$ in Eq. (\ref{CCRs}) gives rise to an equation: 
\begin{eqnarray}
- \tilde g^{0 0} [ \dot \phi, \phi^\prime ] + \xi \phi ( \tilde g^{0 0} g^{\rho\sigma} - \tilde g^{0 \rho} g^{0 \sigma} ) 
[ \dot g_{\rho\sigma}, \phi^\prime ] = - i \delta^3.
\label{pi-phi2}  
\end{eqnarray}
Next, $[ \pi_\phi, g^\prime_{\mu\nu} ] = 0$ produces an equation:
\begin{eqnarray}
- \tilde g^{0 0} [ \dot \phi, g_{\mu\nu}^\prime ] + \xi \phi ( \tilde g^{0 0} g^{\rho\sigma} - \tilde g^{0 \rho} g^{0 \sigma} ) 
[ \dot g_{\rho\sigma}, g_{\mu\nu}^\prime ] = 0.
\label{pi-phi-g}  
\end{eqnarray}
Moreover, $[ {\cal{D}}^{\lambda\rho\sigma} \dot g_{\rho\sigma}, g_{\mu\nu}^\prime ] = 0$, which 
stems from the ${\cal{D}}$-equation (\ref{D-eq}), reads
\begin{eqnarray}
(g^{0 \lambda} g^{\rho\sigma} - 2 g^{\lambda\rho} g^{0 \sigma}) [ \dot g_{\rho\sigma}, g_{\mu\nu}^\prime ] = 0.
\label{D-eq2}  
\end{eqnarray}
Similarly, $[ {\cal{D}}^{\lambda\rho\sigma} \dot g_{\rho\sigma}, \phi^\prime ] = 0$ gives us an equation:
\begin{eqnarray}
(g^{0 \lambda} g^{\rho\sigma} - 2 g^{\lambda\rho} g^{0 \sigma}) [ \dot g_{\rho\sigma}, \phi^\prime ] = 0.
\label{D-eq3}  
\end{eqnarray}

Now we are willing to solve Eqs. (\ref{pi-phi2})-(\ref{D-eq3}). First of all, let us focus on Eq. (\ref{D-eq3}).
From the argument of symmetry, $[ \dot g_{\rho\sigma}, \phi^\prime ]$ must be of form:
\begin{eqnarray}
[ \dot g_{\rho\sigma}, \phi^\prime ] = a_1 ( g_{\rho\sigma} + a_2 \delta_\rho^0 \delta_\sigma^0 ) \delta^3,  
\label{dot-G-Phi}  
\end{eqnarray}
where $a_1, a_2$ are certain coefficients to be determined sooner. Indeed, from Eq. (\ref{D-eq3}), we find that
$a_2 = \frac{2}{g^{00}}$, so the ETCR (\ref{dot-G-Phi}) reads 
\begin{eqnarray}
[ \dot g_{\rho\sigma}, \phi^\prime ] = a_1 \left( g_{\rho\sigma} + \frac{2}{g^{00}} \delta_\rho^0 
\delta_\sigma^0 \right) \delta^3.  
\label{dot-G-Phi2}  
\end{eqnarray}
 
Next, let us solve Eq. (\ref{D-eq2}). In this case, we also find that $[ \dot g_{\rho\sigma}, g_{\mu\nu}^\prime ]$ 
has a symmetry under the simultaneous exchange of $(\mu\nu) \leftrightarrow (\rho\sigma)$ and primed 
$\leftrightarrow$ unprimed in addition to the usual symmetry $\mu \leftrightarrow \nu$ and $\rho \leftrightarrow \sigma$. 
Then, we can write down its general expression like
\begin{eqnarray}
[ \dot g_{\rho\sigma}, g_{\mu\nu}^\prime ] &=& \biggl\{ c_1 g_{\rho\sigma} g_{\mu\nu} + c_2 ( g_{\rho\mu} g_{\sigma\nu}
+ g_{\rho\nu} g_{\sigma\mu} )
\nonumber\\
&+& h \tilde f \Bigl[ c_3 ( \delta_\rho^0 \delta_\sigma^0 g_{\mu\nu} + \delta_\mu^0 \delta_\nu^0 g_{\rho\sigma} )
+ c_4 ( \delta_\rho^0 \delta_\mu^0 g_{\sigma\nu} + \delta_\rho^0 \delta_\nu^0 g_{\sigma\mu} 
\nonumber\\
&+& \delta_\sigma^0 \delta_\mu^0 g_{\rho\nu} + \delta_\sigma^0 \delta_\nu^0 g_{\rho\mu} ) \Bigr]
+ ( h \tilde f )^2 c_5 \delta_\rho^0 \delta_\sigma^0 \delta_\mu^0 \delta_\nu^0 \biggr\} \delta^3,  
\label{dot-G-G}  
\end{eqnarray}
where $c_i ( i = 1, \cdots, 5)$ are some coefficients. Imposing Eq. (\ref{D-eq2}) on (\ref{dot-G-G}) leads to 
relations among the coefficients:
\begin{eqnarray}
c_3 = 2 ( c_1 + c_2 ), \qquad c_4 = - c_2, \qquad c_5 = 4 ( c_1 + c_2 ).
\label{c-relation}  
\end{eqnarray}

Furthermore, using Eq. (\ref{dot-G-Phi2}), Eq. (\ref{pi-phi2}) gives $[ \dot \phi, \phi^\prime ]$,
which is of form:
\begin{eqnarray}
[ \dot \phi, \phi^\prime ] = ( i \tilde f + 3 a_1 \xi \phi ) \delta^3.
\label{dot-Phi-Phi}  
\end{eqnarray}
Finally, with the help of Eqs. (\ref{dot-G-Phi2}), (\ref{dot-G-G}) and (\ref{c-relation}), Eq. (\ref{pi-phi-g}) 
leads to a relation:
\begin{eqnarray}
a_1 = ( 3 c_1 + 2 c_2 ) \xi \phi.
\label{a-c-relation}  
\end{eqnarray}

In order to fix the coefficients $a_1, c_1$ and $c_2$ completely, we need to have two independent relations 
among them. Such the relations can be provided by calculating $[ \dot g_{kl}, g_{mn}^\prime ]$ 
explicitly in terms of $[ \pi_g^{kl}, g_{mn}^\prime  ] = - i \frac{1}{2} ( \delta_m^k \delta_n^l 
+ \delta_m^l \delta_n^k) \delta^3$ in Eq. (\ref{CCRs}) and the concrete expression of $\pi_g^{kl}$ 
in Eq. (\ref{Grav-CCM}). To do that, from Eq. (\ref{Grav-CCM}), let us write
\begin{eqnarray}
\pi_g^{kl} = \hat A^{kl} + \hat B^{kl\rho} b_\rho + \hat C^{klmn} \dot g_{mn} + \hat D^{kl} \dot \varphi.
\label{Pi-G-CCM}  
\end{eqnarray}
Here $\hat A^{kl}, \hat B^{kl\rho}, \hat C^{klmn}$ and $\hat D^{kl}$ commute with $g_{mn}$, and $\hat C^{klmn}$
and $\hat D^{kl}$ are defined as\footnote{It turns out that the concrete expressions of $\hat A^{kl}$ and 
$\hat B^{kl\rho}$ are irrelevant to the calculation of $[ \dot g_{kl}, g_{mn}^\prime ]$.}
\begin{eqnarray}
\hat C^{klmn} = \frac{1}{2} h \varphi K^{klmn}, \qquad
\hat D^{kl} = \tilde g^{00} g^{kl} - \tilde g^{0k} g^{0l},
\label{Pi-G-CCM2}  
\end{eqnarray}
where the definition of $K^{klmn}$ and its property are given by
\begin{eqnarray}
&{}& K^{klmn} = \left|
\begin{array}{rrr}
g^{00} & g^{0l} & g^{0n} \\
g^{k0} & g^{kl} & g^{kn} \\
g^{m0} & g^{ml} & g^{mn} \\
\end{array}
\right|,
\nonumber\\
&{}& K^{klmn} \frac{1}{2} (g^{00})^{-1} ( g_{ij} g_{mn} - g_{im} g_{jn} - g_{in} g_{jm} )
= \frac{1}{2} ( \delta_i^k \delta_j^l + \delta_i^l \delta_j^k ).
\label{Pi-G-CCM3}  
\end{eqnarray}

From Eq. (\ref{Pi-G-CCM}), we can calculate 
\begin{eqnarray}
[ \dot g_{kl}, g_{mn}^\prime ] = \hat C^{-1}_{klpq} \left( [ \pi_g^{pq}, g_{mn}^\prime ] 
-  \hat B^{pq\rho} [ b_\rho, g_{mn}^\prime ] -  \hat D^{pq} [ \dot \varphi, g_{mn}^\prime ] \right).
\label{Pi-G-CCM4}  
\end{eqnarray}
Then, using Eqs. (\ref{CCRs}), (\ref{g-b}), (\ref{dot-G-Phi2}) and (\ref{Pi-G-CCM3}), we find 
\begin{eqnarray}
[ \dot g_{kl}, g_{mn}^\prime ] &=& \tilde f \varphi^{-1} \left[ ( -i - a_1 \tilde g^{00} \xi \phi ) g_{kl} g_{mn} 
+ i ( g_{km} g_{ln} + g_{kn} g_{lm} ) \right] \delta^3
\nonumber\\
&=& \left[ c_1 g_{kl} g_{mn} + c_2 ( g_{km} g_{ln} + g_{kn} g_{lm} ) \right] \delta^3,
\label{Pi-G-CCM5}  
\end{eqnarray}
where the last equality comes from Eq. (\ref{dot-G-G}). In this way, we have suceeded in getting two
independent relations among $a_1, c_1$ and $c_2$:
\begin{eqnarray}
c_1 = \tilde f \varphi^{-1} ( -i - a_1 \tilde g^{00} \xi \phi ), \qquad 
c_2 = i \tilde f \varphi^{-1}.
\label{a-2c}  
\end{eqnarray}

Using Eqs. (\ref{a-c-relation}) and (\ref{a-2c}), we can fix completely the coefficients as
\begin{eqnarray}
c_1 = - i \frac{4 \xi + 1}{6 \xi + 1} \tilde f \varphi^{-1}, \qquad 
c_2 = i \tilde f \varphi^{-1}, \qquad 
a_1 = - i \frac{2}{6 \xi + 1} \tilde f \phi^{-1}.
\label{a-2c-result}  
\end{eqnarray}
Accordingly, we can obtain the following ETCRs:
\begin{eqnarray}
[ \dot g_{\rho\sigma}, \phi^\prime ] = - \frac{2}{6 \xi + 1} i \tilde f \phi^{-1} 
\left( g_{\rho\sigma} + \frac{2}{g^{00}} \delta_\rho^0 \delta_\sigma^0 \right) \delta^3.
\label{ETCR-final1}  
\end{eqnarray}
\begin{eqnarray}
[ \dot \phi, \phi^\prime ] = \frac{1}{6 \xi + 1} i \tilde f \delta^3.  
\label{ETCR-final2}  
\end{eqnarray}
\begin{eqnarray}
[ \dot g_{\rho\sigma}, g_{\mu\nu}^\prime ] &=& i \tilde f \varphi^{-1} 
\biggl\{ - \frac{4 \xi + 1}{6 \xi + 1} g_{\rho\sigma} g_{\mu\nu} + g_{\rho\mu} g_{\sigma\nu}
+ g_{\rho\nu} g_{\sigma\mu} 
\nonumber\\
&+& h \tilde f \Bigl[ \frac{4 \xi}{6 \xi + 1} ( \delta_\rho^0 \delta_\sigma^0 g_{\mu\nu} 
+ \delta_\mu^0 \delta_\nu^0 g_{\rho\sigma} )
- ( \delta_\rho^0 \delta_\mu^0 g_{\sigma\nu} + \delta_\rho^0 \delta_\nu^0 g_{\sigma\mu} 
\nonumber\\
&+& \delta_\sigma^0 \delta_\mu^0 g_{\rho\nu} + \delta_\sigma^0 \delta_\nu^0 g_{\rho\mu} ) \Bigr]
+ ( h \tilde f )^2 \frac{8 \xi}{6 \xi + 1} \delta_\rho^0 \delta_\sigma^0 \delta_\mu^0 \delta_\nu^0 \biggr\} 
\delta^3.
\label{ETCR-final3}  
\end{eqnarray}
It is worthwhile to notice that these ETCRs have two peculiar features, one of which is the presence 
of the factor $6 \xi + 1$ in the denominator, thereby implying that they do not make sense in a theory 
with a local scale (or Weyl) symmetry. In other words, in the case of $6 \xi + 1 = 0$ corresponding to
the Weyl invariant scalar-tensor gravity, we need to introduce one more gauge condition such as $R = 0$ 
or $\phi = {\rm{constant}}$ to fix the Weyl symmetry.   
The other important feature is the existence of the field $\phi$ (or $\varphi$) in the denominator, which
means that an unbroken phase $\langle \phi(x) \rangle = 0$ cannot be dealt with in the present formalism.
This fact strongly suggests that a (global) scale invariance must be broken spontaneously even in quantum
gravity as in classical gravity in order to construct a consistent quantum theory of the scale invariant 
scalar-tensor gravity.

\section{Spontaneous symmetry breakdown of scale invariance}

In the previous work, we have shown that the scale invariance is in fact spontaneously broken 
in classical gravity where the gravitational sector is not quantized \cite{Oda-R}-\cite{Oda-RWS}. 
In this section, we wish to investigate whether a (global) scale invariance is spontaneously broken even 
in quantum gravity or not.  

Taking $\Omega$ in Eq. (\ref{Res-Weyl}) to be a constant, we can define a scale transformation as 
\begin{eqnarray}
g_{\mu\nu} &\rightarrow& g^\prime_{\mu\nu} = \Omega^2 g_{\mu\nu}, \qquad 
\phi \rightarrow \phi^\prime = \Omega^{-1} \phi,
\nonumber\\
b_\rho &\rightarrow& b^\prime_\rho = \Omega^{-2} b_\rho, \qquad
\bar c_\rho \rightarrow \bar c^\prime_\rho = \Omega^{-1} \bar c_\rho, \qquad
\bar c^\rho \rightarrow \bar c^{\rho \prime} = \Omega^{-1} \bar c^\rho, 
\label{G-scale}  
\end{eqnarray}
where we have added the scale transformation for the Nakanishi-Lautrup field and the FP (anti-)ghosts.
Then, it is easy to see that the quantum Langrangian (\ref{ST-q-Lag}) is invariant under the scale 
transformation (\ref{G-scale}). This fact implies that the de Donder gauge is invariant under the 
scale transformation. Incidentally, the de Donder gauge is not invariant under a local scale (or Weyl)
transformation. To make a gauge condition for the general coordinate invariance be invariant under 
the Weyl transformation requires us to take a different gauge condition such as 
$\partial_\mu ( ( -g )^{\frac{1}{4}} g^{\mu\nu} ) = 0$ or $\partial_\mu ( \sqrt{-g} \phi^2 g^{\mu\nu} ) = 0$ 
from the de Donder gauge.

Since the scale transformation is a global one, we can construct a conserved Noether current and charge
along the standard procedure. After some calculations, it turns out that the conserved current for the scale 
symmetry is given by\footnote{See Appendix B.} 
\begin{eqnarray}
J^\mu = \tilde g^{\mu\nu} \left[ \frac{6 \xi + 1}{2}  \partial_\nu ( \phi^2 ) + 2 b_\nu + i \partial_\nu 
( \bar c_\rho c^\rho ) \right].  
\label{Scale-current}  
\end{eqnarray}
It is straightforward to verify that this current is conserved, $\partial_\mu J^\mu = 0$, in terms of 
field equations. Note that in the conformal coupling $\xi = - \frac{1}{6}$, the first term on the RHS 
coming from the classical action is identically vanishing \cite{Jackiw, Oda-U} while the second and third terms coming from 
the gauge-fixing and FP ghost terms, respectively, do not so.  

The Noether charge $Q = \int d^3 x J^0$ turns out to generate the infinitesimal scale transformation correctly 
by using the ETCRs obtained thus far: 
\begin{eqnarray}
\delta g_{\mu\nu} &=& 2 \Lambda g_{\mu\nu}, \qquad
\delta \phi = - \Lambda \phi, \qquad
\delta b_\rho = - 2 \Lambda b_\rho,
\nonumber\\
\delta \bar c_\rho &=& - \Lambda \bar c_\rho, \qquad
\delta c^\rho = - \Lambda c^\rho,
\label{Inf-Scale-Transf}  
\end{eqnarray}
where we have set $\Omega = e^\Lambda \approx 1 + \Lambda$. It might be curious about why no derivative
of the metric tensor $g_{\mu\nu}$ appears in $Q$ since it usually generates the transformation of the metric
tensor \cite{Fujii}. This problem can be understood for the first time in the present formulation since
we have successfully quantized the metric field and the scalar field. 

To check that the charge $Q$ indeed generates the scale transformation (\ref{Inf-Scale-Transf}), let us 
calculate $\delta g_{\mu\nu}$ and $\delta b_\rho$ explicitly. As for $\delta g_{\mu\nu}$, 
\begin{eqnarray}
\delta g_{\mu\nu} &\equiv& [ i \Lambda Q, g_{\mu\nu} ]
\nonumber\\
&=& i \Lambda \int d^3 x^\prime \left[ \tilde g^{0\rho\prime} \left( \frac{6 \xi + 1}{2}  \partial_\rho ( \phi^{\prime2} ) 
+ 2 b_\rho^\prime + i \partial_\rho ( \bar c_\sigma^\prime c^{\sigma\prime} ) \right), g_{\mu\nu} \right]
\nonumber\\
&=& i \Lambda \int d^3 x^\prime \Biggl \{ ( 6 \xi + 1 ) \tilde g^{00\prime} \phi^\prime [ \dot \phi^\prime, g_{\mu\nu} ] 
+ 2 \tilde g^{0\rho\prime} [ b_\rho^\prime, g_{\mu\nu} ] \Biggr \}
\nonumber\\
&=& 2 \Lambda g_{\mu\nu},
\label{g-Inf-Scale-Transf}  
\end{eqnarray}
where in the third equality we put $x^0 = x^{\prime 0}$, and used Eqs. (\ref{g-b}) and (\ref{ETCR-final1}). 
Against the expectation that the derivative of $\phi$ would play a role \cite{Fujii}, the $b$ field also does the job 
in generating the scale transformation. In a similar manner, as for $\delta b_\rho$, we have
\begin{eqnarray}
\delta b_\rho &\equiv& [ i \Lambda Q, b_\rho ]
\nonumber\\
&=& i \Lambda \int d^3 x^\prime \left[ \tilde g^{0\nu\prime} \left( \frac{6 \xi + 1}{2}  \partial_\nu ( \phi^{\prime2} ) 
+ 2 b_\nu^\prime + i \partial_\nu ( \bar c_\sigma^\prime c^{\sigma\prime} ) \right), b_\rho \right]
\nonumber\\
&=& i \Lambda \int d^3 x^\prime \Biggl \{ [ \tilde g^{0\nu\prime}, b_\rho ] \left( \frac{6 \xi + 1}{2}  \partial_\nu 
( \phi^{\prime2} ) + 2 b_\nu^\prime + i \partial_\nu ( \bar c_\sigma^\prime c^{\sigma\prime} ) \right)
\nonumber\\
&+& \tilde g^{00\prime} \Biggl( ( 6 \xi + 1 ) \phi^\prime [ \dot \phi^\prime, b_\rho ] 
+ i \biggl( [ \dot{\bar c}_\sigma^\prime, b_\rho ] c^{\sigma\prime} + \bar c_\sigma^\prime
[ \dot c^{\sigma\prime}, b_\rho ]  \biggr) \Biggr) \Biggr \}
\nonumber\\
&=& - 2 \Lambda b_\rho,
\label{b-Inf-Scale-Transf}  
\end{eqnarray}
where we have used Eqs. (\ref{3-g-b}), (\ref{c-b}) and (\ref{dot-phi-b2}). We wish to mention again 
that in case of classical gravity, one cannot show that the charge $Q$ generates the scale transformation 
owing to the absence of the ETCRs relevant to the Nakanishi-Lautrup field $b_\rho$ and the FP ghosts 
$\bar c_\sigma$ and $c^\sigma$.

Now let us move to the issue of spontaneous symmetry breakdown of scale symmetry in quantum gravity. 
From Eq. (\ref{Inf-Scale-Transf}) and the definition of $\delta \Phi \equiv [ i \Lambda Q, \Phi ]$
for a generic field $\Phi$, we obtain that
\begin{eqnarray}
&{}&[ i Q, g_{\mu\nu} ] = 2 g_{\mu\nu}, \qquad
[ i Q, \phi ] = - \phi, \qquad
[ i Q, b_\rho ] = - 2 b_\rho,
\nonumber\\
&{}&[ i Q, \bar c_\rho ] = - \bar c_\rho, \qquad
[ i Q, c^\rho ] = - c^\rho.
\label{Inf-Scale-Transf2}  
\end{eqnarray}
Assuming that the fields take the following vacuum expectation values: 
\begin{eqnarray}
&{}&\langle 0 | g_{\mu\nu} | 0 \rangle = \eta_{\mu\nu}, \qquad
\langle 0 | \phi | 0 \rangle = \phi_0, \qquad
\langle 0 | b_\rho | 0 \rangle = 0,
\nonumber\\
&{}&\langle 0 | \bar c_\rho | 0 \rangle = 0, \qquad
\langle 0 | c^\rho | 0 \rangle = 0,
\label{Assump-VEV}  
\end{eqnarray}
with $\phi_0$ being a constant,  Eq. (\ref{Inf-Scale-Transf2}) implies that
\begin{eqnarray}
&{}& \langle 0 |[ i Q, g_{\mu\nu} ] | 0 \rangle = 2 \eta_{\mu\nu}, \qquad
\langle 0 | [ i Q, \phi ] | 0 \rangle = - \phi_0, \qquad
\langle 0 |[ i Q, b_\rho ] | 0 \rangle = 0,
\nonumber\\
&{}&\langle 0 | [ i Q, \bar c_\rho ] | 0 \rangle = 0, \qquad
\langle 0 | [ i Q, c^\rho ] | 0 \rangle = 0.
\label{Field-VEV}  
\end{eqnarray}
The second equation in Eq. (\ref{Field-VEV}) shows that the scale invariance is spontaneously
broken at the quantum level as long as $\phi_0 \neq 0$ holds \cite{Oda-R, Oda-RWS}. 

Here three important remarks are in order. First, the first assumption in Eq. (\ref{Assump-VEV}),
$\langle 0 | g_{\mu\nu} | 0 \rangle = \eta_{\mu\nu}$, comes from our postulate that the vacuum
is invariant under translation \cite{N-O-text}:
\begin{eqnarray}
P_\mu | 0 \rangle = 0,
\label{Vac-translation}  
\end{eqnarray}
which means that translational invariance is not broken spontaneously. Moreover, with this assumption 
the graviton can be identified with a Nambu-Goldstone (NG) boson corresponding to spontaneous symmetry 
breakdown of $GL(4)$ symmetry down to $SO(1, 3)$ Lorentz symmetry, thereby proving the exact masslessness 
of the graviton \cite{NO}.

Second, let us pay our attention to the second assumption in Eq. (\ref{Assump-VEV}),
$\langle 0 | \phi | 0 \rangle = \phi_0 \neq 0$, which is also physically plausible by the following argument:
As usual, let us consider to move from the Jordan frame to the Einstein frame by implementing
a local scale transformation only for the metric tensor field as
\begin{eqnarray}
g_{\mu\nu} \rightarrow g_{\ast\mu\nu} = \Omega(x)^2 g_{\mu\nu} 
= \frac{1}{M_{Pl}^2} \xi \phi^2 g_{\mu\nu},
\label{Transf-J-E}  
\end{eqnarray}
where $M_{Pl}$ is the reduced Planck mass. Then, in the Einstein frame, up to a surface term
the quantum Lagrangian (\ref{ST-q-Lag}) is reduced to the form:
\begin{eqnarray}
{\cal{L}}_q = \sqrt{-g_\ast} \left( \frac{M_{Pl}^2}{2} R_\ast - \frac{1}{2} g_\ast^{\mu\nu} \partial_\mu \sigma 
\partial_\nu \sigma \right)
- e^{- \frac{2 \zeta}{M_{Pl}} \sigma} \left( \tilde g_\ast^{\mu\nu} \partial_\mu b_\nu 
- i \tilde g_\ast^{\mu\nu} \partial_\mu \bar c_\rho \partial_\nu c^\rho \right),
\label{E-frame-Lag}  
\end{eqnarray}
where we have defined a scalar field $\sigma(x)$ and a constant $\zeta$ as 
\begin{eqnarray}
\phi = \xi^{- \frac{1}{2}} M_{Pl} e^{\frac{\zeta}{M_{Pl}} \sigma}, \qquad 
\zeta = \sqrt{\frac{\xi}{6 \xi + 1}}.  
\label{Sigma-Zeta}  
\end{eqnarray}
In this way, we can arrive at the Lagrangian in the Einstein frame by starting with that in the Jordan frame.
The key point for the change of the frames lies in Eq. (\ref{Transf-J-E}) where the scale factor $\Omega(x)$
is proportional to $\frac{1}{M_{Pl}} \phi (x)$. Namely, the existence of the non-vanishing ``dilaton'' $\phi \neq 0$, 
or more precisely, $\langle 0 | \phi | 0 \rangle \neq 0$, makes it possible to move from the Jordan frame to
the Einstein one. In this sense, our assumption $\langle 0 | \phi | 0 \rangle = \phi_0 \neq 0$ makes
sense physically.

As a final remark, as mentioned in the previous section, in order to make a consistent quantum gravity based on
the scale invariant scalar-tensor gravity, it is necessary to require the condition $\langle 0 | \phi | 0 \rangle \neq 0$.  
Any physical theories must be formulated within the framework of quantum field theories, so it is natural to
assume such the condition.

To close this section, let us verify more explicitly that the spontaneous symmetry breakdown of the scale symmetry 
occurs in the Einstein frame where the scale symmetry is replaced with a shift symmetry. For this purpose, let us 
rewrite the charge for the scale symmetry in the Jordan frame into that in the Einstein frame as
\begin{eqnarray}
Q = \int d^3 x \, \tilde g_\ast^{0\nu} \left[ \frac{M_{Pl}}{\zeta} \partial_\nu \sigma 
+ e^{- \frac{2 \zeta}{M_{Pl}} \sigma} \left( 2 b_\nu + \partial_\nu ( \bar c_\rho c^\rho ) \right) \right], 
\label{Shift-charge}  
\end{eqnarray}
where we have used Eqs. (\ref{Scale-current}), (\ref{Transf-J-E}) and (\ref{Sigma-Zeta}).
Since $Q$ has a linear term in $\sigma(x)$, the charge cannot annihilate the vacuum $| 0 \rangle$:
\begin{eqnarray}
Q | 0 \rangle \neq 0,
\label{Q-vac}  
\end{eqnarray}
which means the spontaneous symmetry breakdown of scale symmetry in the Jordan frame or shift symmetry
in the Einstein frame.
Actually, from the Lagrangian (\ref{E-frame-Lag}) the canonical conjugate momentum for the scalar field 
$\sigma(x)$ reads
\begin{eqnarray}
\pi_\sigma \equiv \frac{\partial {\cal{L}}}{\partial \partial_0 \sigma} = - \sqrt{-g_\ast} g_\ast^{0\nu} 
\partial_\nu \sigma.
\label{sigma-CCM}  
\end{eqnarray} 
Then, $Q$ can be rewritten as 
\begin{eqnarray}
Q = - \frac{M_{Pl}}{\zeta}  \int d^3 x \, \pi_\sigma + \cdots,
\label{Noether charge-pi}  
\end{eqnarray} 
where $\dots$ denote contributions from the Nakanishi-Lautrup field and the FP ghosts.
Using the equal-time commutation relation $[ \sigma, \pi_\sigma^\prime ] = i \delta^3$,
we obtain
\begin{eqnarray}
[ i Q, \sigma(x) ] = - \frac{M_{Pl}}{\zeta}.
\label{Q-sigma-CR}  
\end{eqnarray} 
Taking the vacuum expectation value of this equation yields
\begin{eqnarray}
\langle 0 | [ i Q, \sigma(x) ] | 0 \rangle = - \frac{M_{Pl}}{\zeta} \neq 0,
\label{<Q>}  
\end{eqnarray} 
which clearly means the spontaneous symmetry breakdown and that the scalar field $\sigma(x)$ is 
the massless NG boson for the shift symmetry.

\section{Conclusions}

In this article, we have performed a manifestly covariant quantization of a scale invariant gravity which 
is equivalent to the Brans-Dicke gravity \cite{Brans} via the field redefinition of a scalar field. Many of studies of
the Brans-Dicke gravity have been thus far confined to the classical analysis, so it is expected that our quantum 
formulation of the Brans-Dicke gravity could provide us with some useful information on quantum aspects
of the theory. 

Actually, we have presented two new results, one of which is that in classically scale invariant gravitational theories,
we have a quantum scale symmetry in addition to a huge choral symmetry when we choose the de Donder gauge 
for the general coordinate invariance. In this respect, it is worthwhile to recall that in the manifestly scale invariant 
regularization method \cite{Englert}-\cite{Ghilencea}, the scale invariance is free of scale anomaly. 

As the second result, we have shown that the scale symmetry is spontaneously broken by quantum effects, 
thus proving that the ``dilaton'' is exactly massless because of the Nambu-Goldstone theorem. 
As pointed out in the previous paper \cite{Oda-SI}, when the dilaton is exactly massless at the quantum level, 
it mediates a long-range force between massive objects as in the Newtonian force, which imposes a severe phenomenological 
constraint on parameters in the Brans-Dicke model \cite{Brans} since the long-range force stemming from the massless 
dilaton could affect the perihelion advance of Mercury, for instance.
  
Furthermore, we have shown that the choral symmetry, which is a Poincar${\rm{\acute{e}}}$-like $IOSp(8|8)$ 
supersymmetry, can be derived from any gravitational theories, which are invariant under the general coordinate
transformation (GCT), if the GCT is gauge-fixed by the de Donder gauge. To put it differently, the choral symmetry
comes from only the de Donder gauge for the GCT in the BRST formalism. 

We still have a lot of works to be done in future. For instance, we would like to extend the present formalism
to gravitational theories with a local scale invariance (or Weyl invariance) and investigate the resultant residual
symmetries. In the case of the Weyl invariance, it seems that we might prefer the Weyl-invariant gauge conditions
such as $\partial_\mu ( ( -g )^{\frac{1}{4}} g^{\mu\nu} ) = 0$ and $\partial_\mu ( \sqrt{-g} \phi^2 g^{\mu\nu} ) = 0$
to the de Donder gauge condition $\partial_\mu ( \sqrt{-g} g^{\mu\nu} ) = 0$ since we have the restricted
Weyl symmetry in these Weyl invariant gauge conditions when we take the gauge condition $R = 0$ for the Weyl invariance. 
However, then we will lose the choral symmetry but instead we would have new residual symmetries. 

As another problem, it is of interest to construct a manifestly scale-invariant regularization scheme in the
theory at hand and calculate an effective potential explicitly \cite{Oda-C}-\cite{Oda-P}. With this regularization scheme,
it is necessary to introduce an additional scalar field in addition to the dilaton and the two scalar fields might collaborate with
each other for nullifying the scalar force. In fact, such an approach on the basis of the dilaton and the axion has been
recently proposed \cite{Burgess}.  In near future, we would like to report these problems.

\begin{flushleft}
{\bf Acknowledgements}
\end{flushleft}

We would like to be grateful to T. Kugo for valuable discussions, in particular, on Sections 2 and 3. 
This work is partly supported by the JSPS Kakenhi Grant No. 21K03539.

\appendix
\addcontentsline{toc}{section}{Appendix~\ref{app:scripts}: Training Scripts}
\section*{Appendix}
\label{app:scripts}
\renewcommand{\theequation}{A.\arabic{equation}}
\setcounter{equation}{0}

\section{Derivation of $[ R_{\mu\nu}, b_\rho^\prime ]$}

In this appendix, we wish to prove the tensorlike ETCR:
\begin{eqnarray}
[ R_{\mu\nu}, b_\rho^\prime ] = - i \tilde f ( \delta_\mu^0 R_{\rho\nu} + \delta_\nu^0 R_{\rho\mu} ) \delta^3. 
\label{R-b2}  
\end{eqnarray} 
Our strategy for the proof is similar to that of \cite{N-O-text}, but is in essense different from it since 
we have been already able to derive the ETCR (\ref{b-dot-b}) without recourse to the Einstein equation. 
In what follows, we will prove the following two equations:
\begin{eqnarray}
[ G^0 \, _\nu, b_\rho^\prime ] = i \tilde f ( \delta_\rho^0 R^0 \, _\nu - \delta_\nu^0 R^0 \, _\rho ) \delta^3, 
\qquad
[ R_{kl}, b_\rho^\prime ] = 0, 
\label{R-G-b}  
\end{eqnarray} 
from which it is easy to see that we can reach our goal (\ref{R-b2}). 

Let us first prove the former equation in (\ref{R-G-b}).   
From the Einstein equation, which is the first equation in (\ref{q-Field-eq}), $G^0 \, _\nu$ is 
\begin{eqnarray}
G^0 \, _\nu = \frac{1}{\varphi} \left[ ( \nabla^0 \nabla_\nu - \delta^0_\nu \Box ) \varphi + \frac{1}{2} T^0 \, _\nu 
+ \frac{1}{2} ( E^0 \, _\nu - \frac{1}{2} \delta^0_\nu E )  \right]. 
\label{G-0-nu}  
\end{eqnarray}
Note that the first term $( \nabla^0 \nabla_\nu - \delta^0_\nu \Box ) \varphi$ contains no $\ddot \varphi$.
Then, $[ G^0 \, _\nu, b_\rho^\prime ]$ reads
\begin{eqnarray}
[ G^0 \, _\nu, b_\rho^\prime ] &=& \frac{1}{\varphi} \left\{ [ ( \nabla^0 \nabla_\nu - \delta^0_\nu \Box ) \varphi, 
b_\rho^\prime ] + \frac{1}{2} [ T^0 \, _\nu, b_\rho^\prime ] 
+ \frac{1}{2} [ E^0 \, _\nu - \frac{1}{2} \delta^0_\nu E, b_\rho^\prime ]  \right\}
\nonumber\\
&\equiv& \frac{1}{\varphi} ( A_1 + A_2 + A_3 ).
\label{G-b-ETCR}  
\end{eqnarray} 

After some calculations, $A_1$ is found to be
\begin{eqnarray}
A_1 \equiv [ ( \nabla^0 \nabla_\nu - \delta^0_\nu \Box ) \varphi, b_\rho^\prime ]
= i \tilde f ( \delta^0_\rho \nabla^0 \nabla_\nu \varphi - \delta^0_\nu \nabla^0 \nabla_\rho \varphi ) \delta^3.
\label{A1}  
\end{eqnarray} 
In order to evaluate $A_2$, it is necessary to calculate $[ T_{\mu\nu}, b_\rho^\prime ]$ whose result
is given by
\begin{eqnarray}
[ T_{\mu\nu}, b_\rho^\prime ] = - i \tilde f ( \delta_\mu^0 T_{\rho\nu} + \delta_\nu^0 T_{\rho\mu} ) \delta^3.
\label{A2-T}  
\end{eqnarray} 
Using this result, $A_2$ becomes
\begin{eqnarray}
A_2 \equiv \frac{1}{2} [ T^0 \, _\nu, b_\rho^\prime ]
= \frac{1}{2} i \tilde f ( \delta^0_\rho T^0 \, _\nu - \delta^0_\nu T^0 \, _\rho ) \delta^3.
\label{A2}  
\end{eqnarray} 
Finally, evaluating $A_3$ requires us to calculate $[ E_{\mu\nu}, b_\rho^\prime ]$, which is found to be
\begin{eqnarray}
[ E_{\mu\nu}, b_\rho^\prime ] = - i \tilde f ( \delta_\mu^0 E_{\rho\nu} + \delta_\nu^0 E_{\rho\mu} ) \delta^3.
\label{A3-E}  
\end{eqnarray} 
Since we can calculate $[ E, b_\rho^\prime ] =0$, $A_3$ reads
\begin{eqnarray}
A_3 \equiv \frac{1}{2} [ E^0 \, _\nu - \frac{1}{2} \delta^0_\nu E, b_\rho^\prime ]
= \frac{1}{2} i \tilde f ( \delta^0_\rho E^0 \, _\nu - \delta^0_\nu E^0 \, _\rho ) \delta^3.
\label{A3}  
\end{eqnarray} 
Adding $A_1, A_2$ and $A_3$ gives rise to the former equation in (\ref{R-G-b})
since from Eq. (\ref{q-Field-eq}) $R^0 \, _\nu$ is described as 
\begin{eqnarray}
R^0 \, _\nu = \frac{1}{\varphi} \left[ ( \nabla^0 \nabla_\nu + \frac{1}{2} \delta^0_\nu \Box ) \varphi 
+ \frac{1}{2} ( T^0 \, _\nu - \frac{1}{2} \delta^0_\nu T ) + \frac{1}{2} E^0 \, _\nu \right]. 
\label{R-0-nu}  
\end{eqnarray}

Next, let us prove the validity of the latter equation in (\ref{R-G-b}).
From the Einstein equation in (\ref{q-Field-eq}), $R_{kl}$ takes the form:
\begin{eqnarray}
R_{kl} = \frac{1}{\varphi} \left[ ( \nabla_k \nabla_l + \frac{1}{2} g_{kl} \Box ) \varphi 
+ \frac{1}{2} ( T_{kl} - \frac{1}{2} g_{kl} T ) + \frac{1}{2} E_{kl} \right]. 
\label{R-kl}  
\end{eqnarray}
With the help of Eqs. (\ref{A2-T}) and (\ref{A3-E}), we find
\begin{eqnarray}
[ R_{kl}, b_\rho^\prime ] = \frac{1}{\varphi} [ ( \nabla_k \nabla_l + \frac{1}{2} g_{kl} \Box ) \varphi,
b_\rho^\prime ]. 
\label{R-kl-b}  
\end{eqnarray}
Using Eq. (\ref{Gamma-b}) and the formula:
\begin{eqnarray}
[ \ddot \varphi, b_\rho^\prime ] = - 2 i \tilde f \partial_\rho \dot \varphi \delta^3 + 2 i \tilde f  
\partial_\rho \varphi \tilde g^{0k} \partial_k ( \tilde f \delta^3 ),
\label{ddot-varphi-b}  
\end{eqnarray}
which can be proved through (\ref{dot-phi-b2}) and (\ref{dot-phi-dot-b3}), we can show that the RHS
of Eq. (\ref{R-kl-b}) is identically vanishing. Accordingly, we have succeeded in proving our statement (\ref{R-b2}).

\renewcommand{\theequation}{B.\arabic{equation}}
\setcounter{equation}{0}

\section{Conserved current for scale symmetry}

In this appendix, we will present a derivation of the conserved current for the scale symmetry
in Eq. (\ref{Scale-current}) and show that the current is indeed conserved on-shell.

First, let us rewrite the Einstein-Hilbert term in order not to include second order derivatives of the 
metric by the standard technique \cite{Landau}:
\begin{eqnarray}
\sqrt{-g} R = \tilde g^{\mu\nu} ( \partial_\alpha \Gamma^\alpha_{\mu\nu} 
- \partial_\nu \Gamma^\alpha_{\mu\alpha} + \Gamma^\sigma_{\mu\nu} \Gamma^\alpha_{\sigma\alpha} 
-  \Gamma^\sigma_{\mu\alpha} \Gamma^\alpha_{\nu\sigma} ),
\label{EH-term}  
\end{eqnarray}
where the first and second terms can be rewritten as
\begin{eqnarray}
&{}& \tilde g^{\mu\nu} \partial_\alpha \Gamma^\alpha_{\mu\nu} 
= \partial_\alpha ( \tilde g^{\mu\nu} \Gamma^\alpha_{\mu\nu} ) - \partial_\alpha \tilde g^{\mu\nu} \cdot \Gamma^\alpha_{\mu\nu}
= \partial_\alpha ( \tilde g^{\mu\nu} \Gamma^\alpha_{\mu\nu} ) - ( \Gamma^\rho_{\alpha\rho} \tilde g^{\mu\nu} 
- 2 \Gamma^\mu_{\alpha\rho} \tilde g^{\nu\rho} ) \Gamma^\alpha_{\mu\nu}
\nonumber\\
&{}& \tilde g^{\mu\nu} \partial_\nu \Gamma^\alpha_{\mu\alpha} 
= \partial_\nu ( \tilde g^{\mu\nu} \Gamma^\alpha_{\mu\alpha} ) - \partial_\nu \tilde g^{\mu\nu} \cdot \Gamma^\alpha_{\mu\alpha} 
= \partial_\nu ( \tilde g^{\mu\nu} \Gamma^\alpha_{\mu\alpha} ) + \Gamma^\mu_{\rho\sigma} 
\tilde g^{\rho\sigma} \Gamma^\alpha_{\mu\alpha}. 
\label{EH-term-1&2}  
\end{eqnarray}
From Eqs. (\ref{EH-term}) and (\ref{EH-term-1&2}), we have
\begin{eqnarray}
\sqrt{-g} R = \partial_\alpha ( \tilde g^{\mu\nu} \Gamma^\alpha_{\mu\nu} - \tilde g^{\mu\alpha} \Gamma^\beta_{\mu\beta} )
+ \tilde g^{\mu\nu} ( \Gamma^\sigma_{\mu\alpha} \Gamma^\alpha_{\nu\sigma} - \Gamma^\sigma_{\mu\nu} 
\Gamma^\alpha_{\sigma\alpha} ).
\label{EH-term2}  
\end{eqnarray}

Using the formula (\ref{EH-term2}), up to a surface term the part of the scalar-tensor gravity in the classical 
Lagrangian (\ref{BD-Lag}) can be cast to the form:
\begin{eqnarray}
\sqrt{-g} \varphi R = - \sqrt{-g} ( G^\alpha \partial_\alpha \varphi + G \varphi ),
\label{ST-gravity-part}  
\end{eqnarray}
where $G^\alpha$ and $G$ are defined by
\begin{eqnarray}
G^\alpha \equiv g^{\mu\nu} \Gamma^\alpha_{\mu\nu} - g^{\mu\alpha} \Gamma^\beta_{\mu\beta},
\qquad
G \equiv g^{\mu\nu} ( \Gamma^\sigma_{\mu\nu} \Gamma^\alpha_{\sigma\alpha} 
- \Gamma^\sigma_{\mu\alpha} \Gamma^\alpha_{\nu\sigma} ).
\label{Two-G}  
\end{eqnarray}
Then, the quantum Lagrangian (\ref{ST-q-Lag}) takes the form:
\begin{eqnarray}
{\cal{L}}_q = - \sqrt{-g} \left( G^\alpha \partial_\alpha \varphi + G \varphi + \frac{1}{2} g^{\mu\nu} 
\partial_\mu \phi \partial_\nu \phi \right)
- \tilde g^{\mu\nu} \partial_\mu b_\nu 
- i \tilde g^{\mu\nu} \partial_\mu \bar c_\rho \partial_\nu c^\rho.
\label{App-q-Lag}  
\end{eqnarray}

With the infinitesimal scale invariance (\ref{Inf-Scale-Transf}), the Noether theorem provides us
with a formula for the conserved current:
\begin{eqnarray}
\Lambda J^\mu = \sum_i \frac{\partial^R {\cal{L}}_q}{\partial \partial_\mu \Phi_i} \delta \Phi_i,  
\label{Form-J}  
\end{eqnarray}
where we use the right-derivative notation and $\Phi_i = \{ g_{\rho\sigma}, \phi, b_\rho, \bar c_\rho,
c^\rho \}$. It is therefore necessary to evaluate each term on the RHS of the following equation:
\begin{eqnarray}
J^\mu = \frac{\partial^R {\cal{L}}_q}{\partial \partial_\mu g_{\rho\sigma}} 2 g_{\rho\sigma}
- \frac{\partial^R {\cal{L}}_q}{\partial \partial_\mu \phi} \phi
- \frac{\partial^R {\cal{L}}_q}{\partial \partial_\mu b_\rho} 2 b_\rho
- \frac{\partial^R {\cal{L}}_q}{\partial \partial_\mu \bar c_\rho} \bar c_\rho
-  \frac{\partial^R {\cal{L}}_q}{\partial \partial_\mu c^\rho} c^\rho. 
\label{Form-J2}  
\end{eqnarray}
To calculate the first term on the RHS, we need the formulae:
\begin{eqnarray}
&{}& \frac{\partial (\sqrt{-g} G^\alpha )}{\partial \partial_\mu g_{\rho\sigma}} = \sqrt{-g} ( g^{\alpha(\rho} g^{\sigma)\mu} 
- g^{\rho\sigma} g^{\mu\alpha} ),
\nonumber\\
&{}& \frac{\partial (\sqrt{-g} G )}{\partial \partial_\mu g_{\rho\sigma}} = \sqrt{-g} \Biggl[ \frac{1}{2}
g^{\rho\sigma} g^{\alpha\beta} \Gamma^\mu_{\alpha\beta} + \Gamma^\alpha_{\alpha\lambda}
\left( g^{\lambda(\rho} g^{\sigma)\mu} - \frac{1}{2} g^{\lambda\mu} g^{\rho\sigma} \right)
\nonumber\\
&+& \frac{1}{2} \left( g^{\lambda(\rho} \partial_\lambda g^{\sigma)\mu} - \frac{1}{2} g^{\lambda\mu} 
\partial_\lambda g^{\rho\sigma} - g^{\rho\alpha} g^{\beta\sigma} \Gamma^\mu_{\alpha\beta} \right) \Biggr],
\label{Two-G-J}  
\end{eqnarray}
where we have used the symmetrization notation, $A_{(\rho} B_{\sigma)} \equiv \frac{1}{2} ( A_\rho B_\sigma
+ A_\sigma B_\rho )$. After a straightforward calculation, we find that the current is given by
\begin{eqnarray}
J^\mu &=& \tilde g^{\mu\nu} \left[ \frac{6 \xi + 1}{\xi}  \partial_\nu \varphi + 2 b_\nu + i \partial_\nu 
( \bar c_\rho c^\rho ) \right]
\nonumber\\
&=& \tilde g^{\mu\nu} \left[ \frac{6 \xi + 1}{2}  \partial_\nu ( \phi^2 ) + 2 b_\nu + i \partial_\nu 
( \bar c_\rho c^\rho ) \right],
\label{Current-J}  
\end{eqnarray}
where we have used $\varphi \equiv \frac{1}{2} \xi \phi^2$. Note that in case of the Weyl invariant scalar-tensor
gravity where $6 \xi + 1 = 0$ is satisfied, the first term, which is a classical contribution, on the RHS in the above
equation identically vanishes \cite{Jackiw, Oda-U}.

Finally, let us check that this current is in fact conserved in terms of the field equations (\ref{q-Field-eq}).
Taking the derivative of the former current in (\ref{Current-J}) leads to 
\begin{eqnarray}
\partial_\mu J^\mu = \sqrt{-g} \left( \frac{6 \xi + 1}{\xi} \Box \varphi + E \right),
\label{Der-Current-J}  
\end{eqnarray}
where we have used the de Donder gauge and the field equations for the FP ghosts. 
It is easy to see that this expression vanishes by using the field equations (\ref{q-Field-eq}).
To do so, let us take the trace of the Einstein equation, i.e., the first equation, in Eq. (\ref{q-Field-eq}) 
whose result reads
\begin{eqnarray}
( 3 \Box - R ) \varphi - \frac{1}{2} T^\mu \, _\mu + \frac{1}{2} E = 0.
\label{Trace-q-Einstein}  
\end{eqnarray}
From the field equation for the scalar field, i.e., the second equation, in Eq. (\ref{q-Field-eq}), we obtain
\begin{eqnarray}
\varphi R = - \frac{1}{2} \phi \Box \phi.
\label{Scalar-q-eq}  
\end{eqnarray}
Moreoever, taking the trace of $T_{\mu\nu}$ in Eq. (\ref{Def-T}) gives rise to
\begin{eqnarray}
T^\mu \, _\mu = - ( \partial_\rho \phi )^2.
\label{Trace-T-tensor}  
\end{eqnarray}
Then, substituting Eqs. (\ref{Scalar-q-eq}) and (\ref{Trace-T-tensor}) into (\ref{Trace-q-Einstein}),
we have
\begin{eqnarray}
3 \Box \varphi + \frac{1}{2} [ \phi \Box \phi + ( \partial_\rho \phi )^2 ]  + \frac{1}{2} E = 0.
\label{Trace-q-Einstein2}  
\end{eqnarray}
Since the definition $\varphi \equiv \frac{1}{2} \xi \phi^2$ gives us an equation:
\begin{eqnarray}
\Box \varphi = \xi [ \phi \Box \phi + ( \partial_\rho \phi )^2 ],
\label{varphi vs phi}  
\end{eqnarray}
inserting this equation to Eq. (\ref{Trace-q-Einstein2}) yields the desired equation:
\begin{eqnarray}
\frac{6 \xi + 1}{\xi} \Box \varphi + E = 0,
\label{Final EQ}  
\end{eqnarray}
by which we can prove the conservation law of the scale current, $\partial_\mu J^\mu = 0$
as seen in Eq. (\ref{Der-Current-J}).


\end{document}